\documentclass[a4paper,11pt]{article}

\usepackage{jheppub} 
\usepackage{graphicx}
\usepackage{amsmath}
\usepackage[utf8]{inputenc}

\usepackage{amssymb}
\usepackage{amsfonts}
\usepackage{amsbsy}
\usepackage{url}
\usepackage{enumerate}
\usepackage{caption}
\usepackage{subcaption}
\usepackage{color}
\usepackage{bm}
\usepackage{bbm}
\usepackage{comment}
\usepackage{multirow,bigdelim}
\usepackage[head1]{DotArrow}
\usepackage[normalem]{ulem}
\usepackage{CJKutf8}
\usepackage{braket}

\makeatletter
\@addtoreset{equation}{section}
\makeatother

\newcommand{\mr}[1]{\mathrm{#1}}

\makeatletter
\gdef\@fpheader{}
\makeatother

\graphicspath{{./}{./figs/}}
\newbox{\ORCIDicon}
\title{
Gravitational wave spectrum from expanding string loops on domain walls: \\
Implication for nanohertz pulsar timing array signals
}

\author[a,b,c]{Yu Hamada\,\href{https://orcid.org/0000-0002-0227-5919}{\usebox{\ORCIDicon}}}
\emailAdd{yu.hamada@desy.de}
\affiliation[a]{Deutsches Elektronen-Synchrotron DESY, Notkestr. 85, 22607 Hamburg, Germany}
\affiliation[b]{KEK Theory Center, Tsukuba 305-0801, Japan}
\affiliation[c]{Research and Education Center for Natural Sciences, Keio University, 4-1-1 Hiyoshi, Yokohama, Kanagawa 223-8521, Japan}

\author[b]{and Wakutaka Nakano\,\href{https://orcid.org/0000-0001-8480-6965}{\usebox{\ORCIDicon}}}
\emailAdd{wnakano@post.kek.jp}

\abstract{
We analytically calculate the spectrum of stochastic gravitational waves (GWs) emitted by expanding string loops on domain walls in the scenario where domain walls decay by nucleation of string loops. By introducing macroscopic parameters characterizing the nucleation of the loops, the stochastic GW spectrum is derived in a way that is independent of the details of particle physics models. In contrast to GWs emitted from bubble collisions of the false vacuum decay, the string loops do radiate GWs even when they are perfectly circular before their collisions, resulting in that more and more contribution to the spectrum comes from the smaller and smaller loops compared to the typical size of the collided loops. Consequently, the spectrum is linearly proportional to the frequency at the high-frequency region, which is peculiar to this GW source. Furthermore, the results are compared with the recent nano-Hertz pulsar timing array signal, as well as the projected sensitivity curves of future gravitational wave observatories.
}

\preprint{KEK-TH-2619, DESY-24-067}

\begin{document}

\maketitle

\section{Introduction}
Gravitational waves (GWs) have attracted much attention as powerful messengers,
offering unprecedented insights into unexplored fields of the universe. 
Since the first direct detection of gravitational waves in 2015 by the Laser Interferometer Gravitational-Wave Observatory~\cite{LIGOScientific:2016aoc,LIGOScientific:2016emj,LIGOScientific:2016vbw}, 
numerous events, such as black hole mergers~\cite{LIGOScientific:2016vlm} and neutron star mergers~\cite{LIGOScientific:2017vwq}, have been observed,
enriching our understanding of astrophysical phenomena.
Besides the astrophysical sources, 
GWs provide various insights into the early universe as well, 
in which high-energy physics is considered to play significant roles.
In particular, they offer the potential to probe exotic processes such as cosmological first-order phase transitions (FOPT)~\cite{Kosowsky:1991ua,Kosowsky:1992vn,Kamionkowski:1993fg,Grojean:2006bp,Caprini:2009fx,Caprini:2015zlo,Caprini:2018mtu,Caprini:2019egz,Athron:2023xlk} and the motion of topological defects~\cite{Vilenkin:1981bx,Accetta:1988bg,Caldwell:1991jj,Vilenkin:2000jqa},
which do not appear in the standard model (SM) of particle physics.
Therefore, GWs may provide smoking gun of new physics beyond the SM.

Among these phenomena, domain walls (DWs) and cosmic strings stand out as intriguing candidates to generate detectable GW signals.
DWs are hypothetical planar objects that arise when a system undergoes a phase transition in the early universe
accompanied with spontaneous symmetry breaking (SSB) of a discrete symmetry. 
On the other hand, cosmic strings are one-dimensional topological defects that can form during SSB of continuous symmetries such as $U(1)$.
These inhomogeneous cosmic structures can lead to the emission of significant amount of GWs.
Moreover, in various particle physics models, there appear hybrid objects consisting of DWs and cosmic strings~\cite{Kibble:1982dd,Vilenkin:1982ks,Everett:1982nm,Preskill:1992ck,Vilenkin:2000jqa,Kawasaki:2013ae,Eto:2023gfn,Maji:2023fba,Lazarides:2023ksx,Fu:2024rsm}, 
which we call hybrid defects in this paper.
The hybrid defects are crucial for the cosmological validity of the DWs
since the existence of the hybrid defects allows the walls to decay
by nucleation of loops of cosmic strings (Fig.~\ref{fig:hybrid}),
which helps to avoid the stringent bound from cosmological observations.
Once the string loops are nucleated on the DWs,
they expand faster and faster by eating the DWs
like the expanding true-vacuum bubble in the FOPT.
The expanding string loops can be quite energetic because their speed approaches the speed of light
and hence they may radiate a significant amount of GWs.
However, the GWs emitted from this process have been less studied so far.
In Ref.~\cite{Dunsky:2021tih}, they considered a similar decay process whereas the contribution from the expanding loops is not considered.
As another decay process, 
DWs can be collapsed by bias terms~\cite{Gleiser:1998na,Preskill:1991kd,Hiramatsu:2013qaa,Saikawa:2017hiv,Kitajima:2023cek},
which is not driven by the string loops.
Therefore, the detailed analysis of the GW spectrum emitted by the decaying DWs with expanding string loops is still missing.

In this paper, we calculate the spectrum of GWs emitted by expanding string loops on DWs analytically 
in the scenario where DWs decay by nucleation of loops of cosmic strings. 
In order to describe the stochastic process of the loop nucleation,
we introduce a setup similar to that developed for the analytic derivation of the GW spectrum in bubble collisions~\cite{Jinno:2016vai,Jinno:2017fby}.
After introducing parameters to characterize the loop nucleation,
the time duration of the nucleation $\beta^{-1}$,
the energy fraction released from the DWs $\alpha$,
the temperature at which the decay of the walls completes $T_*$,
the width of the walls $d_\mr{DW}$,
and the number of the walls within the Hubble patch $N_\mr{walls}$,
we derive the analytic formula of the stochastic GW spectrum in a way that is independent of details of particle physics models.

In contrast to the GW spectrum from bubble collisions in FOPT,
the string loops do radiate GWs even when they are perfectly circular before collisions.
This is because the $O(2)$ symmetry does not prevent GW radiation unlike $O(3)$.
In addition, since the expanding loops remain on the two dimensional plane,\footnote{This is in contrast to false vacuum decay catalyzed by DWs or cosmic strings studied in Refs.~\cite{Blasi:2023rqi,Blasi:2024mtc}, in which the bubbles nucleated on the walls or strings expand in the three-dimensional bulk.}
the UV behavior of the GW spectrum can be different from other conventional sources.
As a result,
the GW spectrum is proportional to $f^3$ with the frequency $f$ in the IR regime,
which can be deduced from the causality requirement~\cite{Caprini:2009fx},
while it is linearly proportional to $f$ in the UV regime (not even suppressed!),
which is peculiar to this GW source.
This spectrum has a UV cutoff corresponding to either the DW width or the initial radius of the nucleated loops,
at which our calculation breaks down.

Recently, the NANOGrav, EPTA, PPTA, and CPTA groups~\cite{NANOGrav:2023gor,EPTA:2023fyk,Xu:2023wog,Reardon:2023gzh} reported
data showing stochastic GW background in the $\mathcal{O}(1-10)$ nHz frequency band.
We will see that this pulsar timing data can be attributed to the predicted GW spectrum from the hybrid defects with appropriate parameters.
We will also show that this GW spectrum has much potential to be probed by future GW observatories.

This paper is organized as follows.
In Sec.~\ref{sec:hybrid}, we briefly review the hybrid defects consisting of DWs and cosmic strings.
In Sec.~\ref{sec:GW}, we derive the GW spectrum from the hybrid defects.
We compare the predicted stochastic GW spectrum with the pulsar timing data and future gravitational wave observatories in Sec.~\ref{sec:present-spectrum}.
The discussion and conclusion are given in Sec.~\ref{sec:conclusion}.
Appendix~\ref{app-quadrupole} is devoted to calculating the quadrupole of the energy density of the expanding string loop and a naive derivation of the GW spectrum.
Appendix~\ref{sec:k_cancel} is devoted to showing the cancellation of the $k^1$ dependence of the spectrum in the IR regime.

\section{Wall-string hybrid defects}
\label{sec:hybrid}
We here give brief reviews on topological defects; DWs, cosmic strings and their hybrid objects.
We also present some simple examples of particle physics models that give such hybrid defects.

\subsection{Domain wall}
\label{subsec:wall}
Let us consider a real scalar field $\phi$ charged under a discrete $\mathbb{Z}_2$ symmetry.
The simplest Lagrangian is given as
\begin{equation}
 \mathcal{L}= \frac{1}{2}(\partial_\mu \phi)^2 - \lambda_\phi(\phi^2-v_\phi^2)^2 \, ,
\end{equation}
where $\phi$ takes a vacuum expectation value (VEV) $v_\phi$ at the vacuum, breaking the $\mathbb{Z}_2$ symmetry spontaneously.
This phase transition allows us to consider a classical solution of the equations of motion (EOM) that is topologically protected:
\begin{equation}
 \phi = v_\phi \tanh \left[\sqrt{2\lambda_{\phi}} v_\phi \, x \right]\, , 
\end{equation}
which connects the two vacua $\phi= \pm v_\phi$ at $x\to \pm \infty$
and describes an excitation localized around $x=0$ (two-dimensional wall in three dimensions),
called the DW.

The DWs always appear whenever a discrete symmetry that is respected in the Lagrangian is spontaneously broken at the vacuum.
This is characterized by the zeroth order homotopy group $\pi_0(\mathcal{M})$ where $\mathcal{M}$ denotes the vacuum manifold (moduli space of the order parameter) of the theory.
In the above example, the vacuum is represented by the two points, $\phi/v_\phi= \pm 1$, leading to $\pi_0(\mathbb{Z}_2) = \mathbb{Z}_2$.
The DW is labeled by the topological $\mathbb{Z}_2$ charge.
On the other hand, if the theory has a more general discrete symmetry, say, $\mathbb{Z}_N$,
then one can consider $N-1$ types of the DWs characterized by $\pi_0(\mathbb{Z}_N) = \mathbb{Z}_N$.

\subsection{Cosmic string}
\label{subsec:string}
Next, let us consider the Abelian-Higgs model
that consists of a complex scalar field $\Phi$ coupled to a $U(1)$ gauge field $A_\mu$.
The Lagrangian is given as
\begin{equation}
 \mathcal{L}= |D_\mu \Phi|^2 - \lambda_\Phi(|\Phi|^2-v_\Phi^2)^2 - \frac{1}{4} F_{\mu\nu}F^{\mu\nu} \, ,
\end{equation}
where $D_\mu = \partial_\mu - ig A_\mu$ is the covariant derivative and $F_{\mu\nu}$ is the field strength.
After $\Phi$ gets a VEV $v_\Phi$,
a string-like configuration called the cosmic string appears:
\begin{equation}
 \Phi = v_\Phi \, e^{i \theta} \,f(r),\quad A_\theta = \frac{1}{g} a(r) \, ,
\end{equation}
where $f$ and $a$ are scalar functions of the two-dimensional radial coordinate $r=\sqrt{x^2+y^2}$ 
satisfying the boundary conditions
\begin{equation}
 f(0)=a(0)=0 \,,  \quad f(\infty) = a(\infty)=1 \, ,
\end{equation}
and their detailed shapes are determined by solving the EOMs.
The phase of $\Phi$ only depends on the spatial angle around the $z$ axes, $\theta=\mathrm{arccos} (x/r)$,
which means that this phase changes from $0$ to $2\pi$ as the spatial angle $\theta$ does,
implying a winding number unity. 
One can see that a string-like excitation is localized around $r=0$ while it approaches the vacuum configuration as $r\to \infty$.

The cosmic string exists whenever the vacuum manifold $\mathcal{M}$ of the theory is not simply connected,
i.e., whenever it has a closed loop that cannot be contracted to a point.
This is equivalent to stating that it has the non-trivial first homotopy group $\pi_1(\mathcal{M}) \neq 0$.
The above example provides $\mathcal{M}\simeq S^1$ because of $\braket{|\Phi|}=v_\Phi$, and the winding number corresponds to a non-trivial element of $\pi_1(S^1)=\mathbb{Z}$.
More generically, the vacuum manifold is not necessarily $S^1$ but can be a more complicated structure, e.g., $SO(3)\simeq SU(2)/\mathbb{Z}_2$, and the concerned symmetries can be global symmetries instead of gauged ones.
Thus, the detailed properties of the strings depend on the models.
When the model experiences further symmetry breaking in IR, 
the vacuum manifold $\mathcal{M}$ can be changed so that the first homotopy group becomes trivial $\pi_1(\mathcal{M})=0$.
In such a case, the winding number is not topologically protected
and hence the strings cannot be isolated objects but must be attached by DWs.

\subsection{Hybrid defects of the wall and string}
\begin{figure}[tbp]
    \centering    
    \includegraphics[width=0.75\textwidth]{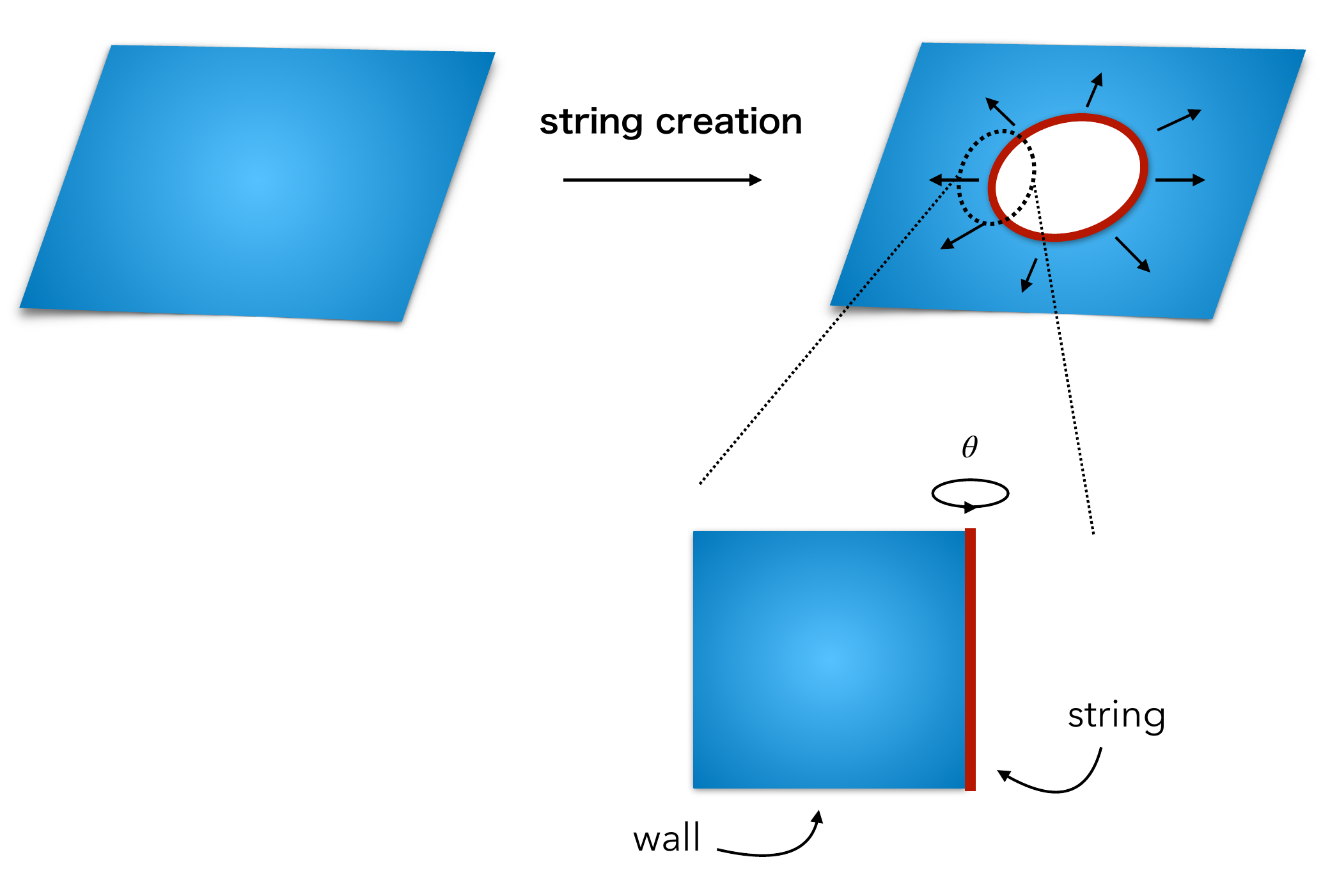}
    \caption{
Schematic of the hybrid object consisting of a DW and cosmic string.
The wall (blue) ends with the loop of the cosmic string (red).
After nucleation of the string loop, it expands by eating the DW.
Locally the string and wall are described by the straight long string in Sec.~\ref{subsec:string} and the flat wall in Sec.~\ref{subsec:wall}, respectively.
    }
    \label{fig:hybrid}
\end{figure}

After creating the DWs, they eventually form a network whose typical size is of the order of the Hubble length.
The DW network easily dominates the energy density of the universe 
because its energy density is inversely proportional to the square of the scale factor $a(t)$ in the radiation-dominated universe, 
which decreases more slowly than that of radiation,
resulting in destruction of the standard cosmology.
One popular way to avoid this is to allow the DWs to decay by nucleation of loops of cosmic strings on them.
These loops are boundaries of empty holes where there is no DW, 
and thus the wall ends at the strings.
After creating the string loop, it can expand faster and faster eating the energy density of the wall
and radiating GWs (see Appendix~\ref{app-quadrupole}),
whose spectrum is investigated in this paper.
The schematic picture of this hybrid object is shown in Fig.~\ref{fig:hybrid}.

This hybrid defect appears in various cases.
For instance, when the universe experiences a two-step SSB~\cite{Kibble:1982dd}, $G \to H \to 1$ with $\pi_0(H) \neq 0$ and $G$ being a simple group,
the first SSB gives rise to the cosmic strings due to $\pi_1(G/H) =  \pi_0(H) \neq 0$\footnote{This comes from the exact sequence of homotopy groups.} and the second one does to the DWs.
Another case is that the model has an approximate $U(1)$ symmetry that is explicitly broken by a tiny breaking parameter in the potential.
This is nothing but the case of axion-like particles.
In both cases, the strings usually appear before the walls and form a network as well.
The string network is here assumed to be diluted away by the cosmic inflation 
because otherwise the string network gets connected by the DWs and eventually decays by the wall tension 
well before the creation of the string loops.
Since the DWs are not topologically stable in the both cases, 
they allow the string loops to be nucleated on them by quantum tunneling or thermal fluctuation,
which are nothing but the hybrid defects stated above.
We here present some concrete examples of particle physics models that can provide hybrid defects of DWs and cosmic strings.
For other studies on hybrid defects of DWs and strings, see Refs.~\cite{Kibble:1982dd,Vilenkin:1982ks,Everett:1982nm,Preskill:1992ck,Vilenkin:2000jqa,Kawasaki:2013ae,Eto:2023gfn}.

\paragraph{Example 1: $SU(2)$ model}
Let us consider two adjoint scalars and one doublet scalar charged under a $SU(2)$ symmetry, 
which can be either of gauge or global.
When the adjoint scalars take VEVs, say, $\Delta^{(1)}=v_\Delta \sigma^1$ and $\Delta^{(2)}=v_\Delta \sigma^2$
($\sigma^a$: the Pauli matrices),
which break the $SU(2)$ symmetry down to the center $\mathbb{Z}_2$ symmetry,
producing cosmic strings
due to $\pi_1(SU(2)/\mathbb{Z}_2)=\mathbb{Z}_2$.
This string is called the $\mathbb{Z}_2$ string~\cite{PhysRevLett.55.2398} because the winding number is characterized by the $\mathbb{Z}_2$ charge.
Due to the cosmic inflation, the produced strings are diluted away.
Afterward, the doublet scalar takes a VEV, so that the center $\mathbb{Z}_2$ symmetry is further broken into the trivial group $1$,
producing DWs.
Since the whole zero-th homotopy group is trivial,
$\pi_0(SU(2))=0$,
these DWs are not topologically stable
but can decay by nucleation of the string loops.

\paragraph{Example 2: Axion-like particle}
It is well known that models with axion-like particles can exhibit wall-string hybrid defects.
Suppose that the model enjoys a global $U(1)$ symmetry, 
which is explicitly broken by a breaking term in the axion-like potential having a single potential minimum.
This breaking term is not effective until some temperature (say, confinement scale of dark QCD)
and thus cosmic strings appear when the $U(1)$ is spontaneously broken, being diluted away by the inflation.
When the breaking term becomes effective or, equivalently, the axion potential is lifted except for the minimum point, 
the DWs are generated like sine-Gordon kinks.
Since there is no discrete symmetry that is spontaneously broken, this DW is not topologically stable but allows the nucleation of string loops.
This is nothing but the case of the axion DW of the DW number unity.

\section{Gravitational waves from wall-string hybrid defects}
\label{sec:GW}
\subsection{Nucleation of string loop}
The GW spectrum is calculated in this section.
Here, we assume the cosmic strings to be produced in advance as stated in the last section.
It gives a condition $v_\mu > v_\sigma$, where $v_{\mu/\sigma}$ represents the VEV of the fields constituting cosmic strings/DWs.
Also, the DWs should not have boundary before the nucleation of the string loops,
which requires the cosmic inflation to occur to dilute the strings away.
Once the string loops are nucleated on DWs, they will expand faster and faster by eating the energy density of DWs~\cite{Dunsky:2021tih}.
In this paper, the runaway case 
is assumed,
i.e., the expansion is not affected by friction.

Each string loop expanding on the DW emits GW even for a flat DW configuration.
This is because the quadrupole moment of the circular expanding loop is non-zero and depends on time,
see Appendix~\ref{app-quadrupole}.
Here, the curvature radius of DW is assumed to be larger than the typical size of the collided loops ($\beta^{-1}$ defined below) 
in order to approximate the DW as a planer object.

Unlike the purely quantum nucleation of the loops in Ref.~\cite{Dunsky:2021tih},
in which the nucleation rate does not depend on time,
we instead consider a tunneling process due to the finite temperature,
which allows us to assume the nucleation rate per unit \textit{area} on the DWs to have the following expression
\begin{align}
    \Gamma (t) \simeq M^3 \exp\left[-\frac{E_\mr{crit.}(\mu,\sigma,T_\mr{temp})}{T_\mr{temp}}\right]
\end{align}
where $M$ is some energy scale and $E_\mr{crit.}$ is the free energy of the nucleated loop on the DW with the critical size $R_c$.
At the zero-temperature, $R_c$ is determined by~\cite{Kibble:1982dd,Preskill:1992ck}
\begin{align}
0 &=  \frac{d}{d R_c} \left( 2\pi R_c \mu - \pi R_c^2 \sigma\right) \\
\therefore \, & E_\mr{crit.}(\mu,\sigma,T_\mr{temp}=0) = \frac{\pi \mu^2}{\sigma}
\end{align}
where $\mu$ and $\sigma$ are the tensions of string and DW.
Taking into account the finite temperature correction,
$\mu$, $\sigma$, and other model parameters depend on $T_\mr{temp}$,
which makes it difficult to obtain the explicit expressions of $E_\mr{crit.}(\mu,\sigma,T_\mr{temp})$ and $\Gamma (t)$.

Inspired from the bubble nucleation process in the FOPT, 
we here simply assume the following expression
\begin{align}
    \Gamma (t) = \Gamma_* e^{\beta (t - t_*)} ,
    \label{eq:cre_rate}
\end{align}
with $t_*$ being some fixed time typically around the decay time and $\Gamma_*$ being the nucleation rate at $t=t_*$.
We have ignored the quadratic and higher-order terms in the exponential.
Ignoring the $T_\mr{temp}$ dependence in the prefactor of $\Gamma(t)$, the parameter $\beta$ is defined as
\begin{align}
\beta = \partial_t\left.\frac{-E (\mu,\sigma, T_\mr{temp})}{T_\mr{temp}}\right|_{t = t_*} 
\end{align}
and corresponds to the inverse of the duration of the nucleation process.
Since it depends on microscopic parameters,
it is calculable by performing the finite-temperature calculation in field theories
once fixing a particle physics model.
Nevertheless, it is beyond the scope of this paper, and hence we do not specify it here but leave it as a macroscopic free parameter.

In the following sections, we implicitly assume a relation\footnote{
This relation is satisfied for the thermal first-order phase transition of bubble nucleation~\cite{Kamionkowski:1993fg}.}
 $d_{\rm DW}^{-1} \gg \beta \gg H_*$, 
where $d_{\rm DW}$ is the width of the DWs and $H_*$ is the Hubble expansion rate at $t=t_*$.
Since $\beta^{-1}$ is regarded as the typical size of the nucleation process,
this relation is stating a reasonable assumption that it should be larger than the DW width and should not exceed the Hubble length.
There is another necessary condition to perform our calculation.
In order for the initial radius of the nucleated loops to be neglected,
the relation $\beta^{-1} \gg R_c$ must be satisfied.
When the finite-temperature correction is subleading, i.e., $R_c \simeq \mu /\sigma$, it is rewritten as
\begin{equation}
\beta^{-1} \gg \frac{\mu}{\sigma} \, ,
\end{equation}
which is typically satisfied in most cases.
If one takes $\mu \simeq v_\mu^2$ and $\sigma\simeq v_\sigma^3$,
the inequality leads to $\beta \ll v_\sigma^3/v_\mu^2$,
which is easily satisfied unless there is huge hierarchy between $v_\sigma$ and $v_\mu$.

\subsection{Abundance of gravitational wave}
Since we expect the production of GWs to complete in a short period compared to the Hubble time,
the background metric is well approximated by the Minkowski one + linearized tensor perturbation,
\begin{align}
    ds^2 = -dt^2 + (\delta_{ij} +2h_{ij}) dx^i dx^j + \mathcal{O}(h^2) \, .
    \label{eq:metric}
\end{align}
Also, the transverse-traceless gauge (TT gauge) is taken in the following calculation,
$h_{ii}=0,\, \partial_j h_{ij}=0$.
Our calculation refers to Ref.~\cite{Jinno:2016vai}. (See also Refs.~\cite{Caprini:2007xq,Jinno:2017fby}.)

The gravitational wave is calculated from the linearized Einstein equation 
\begin{align}
    \ddot{h}^{\rm TT}_{ij} (t,\Vec{k}) +\Vec{k}^2 h^{\rm TT}_{ij} (t,\Vec{k}) = 8\pi G \Pi^{\rm TT}_{ij} (t,\Vec{k}),
\end{align}
where $\Pi^{\rm TT}_{ij} (t,\Vec{k}) = K_{ijkl} (\hat{k}) \Pi_{kl} (t,\Vec{k})$ is the energy momentum tensor after the TT projection 
\begin{align}
    K_{ijkl} (\hat{k}) = & P_{ik} (\hat{k}) P_{jl} (\hat{k}) - \frac{1}{2} P_{ij} (\hat{k}) P_{kl} (\hat{k}) , \\
    P_{ij} (\hat{k}) = & \delta_{ij} - \hat{k}_i \hat{k}_j ,
\end{align}
and $\Pi_{kl} (t,\Vec{k})$ is the energy momentum tensor in the $k$-space.
Here, hat represents the unit vector and the convention of Fourier transformation is taken as $\int d^3 x e^{i \Vec{k} \cdot \Vec{x}}$ and $\int d^3 k /(2\pi)^3 e^{-i \Vec{k} \cdot \Vec{x}}$.

The Einstein equation is solved by using the method of Green's function $G_k (t,t')$
\begin{align}
    h^{\rm TT}_{ij} (t,\Vec{k}) = 8\pi G \int_{-\infty}^{t} dt' G_k (t, t') \Pi^{\rm TT}_{ij} (t', \Vec{k}) ,
    \label{eq:Green}
\end{align}
and the solution is $G_k (t, t') = \sin{(k(t-t'))}/k$.
The gravitational wave propagates as a plane wave after its sources disappear at $t=t_m$.
Then, the coefficients of plane gravitational wave are obtained by matching at time $t_m$ as
\begin{align}
    \label{eq:h_ij}
    h^{\rm TT}_{ij} (t, \Vec{k}) = & A_{ij} (\Vec{k}) \sin{(k(t-t_m))} + B_{ij} (\Vec{k}) \cos{(k(t-t_m))} , \\
    A_{ij} (\Vec{k}) = & 8\pi G \int_{-\infty}^{t_m} dt' \frac{\cos{(k(t_m - t'))}}{k} \Pi^{\rm TT}_{ij} (t', \Vec{k}) , \\
    B_{ij} (\Vec{k}) = & 8\pi G \int_{-\infty}^{t_m} dt' \frac{\sin{(k(t_m - t'))}}{k} \Pi^{\rm TT}_{ij} (t', \Vec{k}) .
\end{align}

The total energy density of the gravitational wave is taken as
\begin{align}
    \rho_{\rm GW} (t) = & \frac{1}{8\pi G} \langle \dot{h}^{\rm TT}_{ij} (t,\Vec{x}) \, \dot{h}^{{\rm TT},ij} (t,\Vec{x}) \rangle_T , 
\end{align}
where $\langle \cdots \rangle_T$ represents the average over ensemble and time during a period of oscillation frequency.

Note that this ensemble average consists of an average over many nucleated loops and a spatial average of DW configurations.\footnote{
This configuration average is reasonable because the present Hubble scale is much larger than that at $t\sim t_*$, $H_0 \gg 10^2 H_*$.}
The equal-time correlator in the $k$-space 
becomes
\begin{align}
    \langle \dot{h}^{\rm TT}_{ij} (t,\Vec{k}_1) \, \dot{h}^{*{\rm TT},ij} (t,\Vec{k}_2) \rangle_T = & {(8\pi G)^2}{}  \int_{-\infty}^{\infty} dt_1  \int_{-\infty}^{\infty} dt_2 \langle \sum_{\{a,b\}}^{\rm walls} \Pi^{{\rm TT}}_{ij,a}(t_1,\Vec{k}_1) \Pi^{*{\rm TT},ij}_{b} (t_2,\Vec{k}_2)\rangle_{\rm ens} \nonumber \\
 & \hspace{3em} \times  \int_{-T_{\rm osc}/2}^{T_{\rm osc}/2} \frac{dt''}{T_{\rm osc}} \cos (k_1(t''-t_1))\cos (k_2(t''-t_2)) ,
\end{align}
where the time average over the oscillation frequency $T_{\rm osc}$ is shown explicitly and $\langle \cdots \rangle_{\rm ens}$ represents the ensemble average.
The indices $a,b$ are introduced to label the DW on which the GW source (string loop) lies.
Here, the matching time is taken to $t_m \to \infty$ approximately because of the expression of nucleation rate Eq.~(\ref{eq:cre_rate}).
If the correlations between different DWs are negligible, 
the energy-momentum tensors inside the correlation function must lie on the same DW, i.e., $a=b$.
Moreover, the spatial average gives the translational invariance 
leading to a delta function $\delta^3(\Vec{k}_1-\Vec{k}_2)$.
In the end, the equal-time correlator becomes
\begin{align}
    \langle \dot{h}^{\rm TT}_{ij} (t,\Vec{k}_1) \, \dot{h}^{*{\rm TT},ij} (t,\Vec{k}_2) \rangle_T = & \frac{(8\pi G)^2}{2}  \int_{-\infty}^{\infty} dt_1  \int_{-\infty}^{\infty} dt_2 \cos (k_1(t_1-t_2)) \nonumber \\
 & \hspace{3em} \times (2\pi)^3 \delta^3(\Vec{k}_1-\Vec{k}_2) \Pi(t_1,t_2,k) ,
\end{align}
where
\begin{align}
    (2\pi)^3 \delta^3(\Vec{k}_1-\Vec{k}_2) \Pi(t_1,t_2,k) =  \langle \sum_{a}^{\rm walls} \Pi^{\rm TT}_{ij,a}(t_1,\Vec{k}_1) \Pi^{*{\rm TT},ij}_{a} (t_2,\Vec{k}_2)\rangle_{\rm ens} \, .
\end{align}
Then, the energy density is rewritten as
\begin{align}
    \rho_{\rm GW} (t) = & \frac{2G}{\pi}\int d(\ln k) \int_{-\infty}^{\infty} dt_1 \int_{-\infty}^{\infty} dt_2 k^3 \cos{(k(t_1 - t_2))}  \Pi(t_1,t_2,k) ,
\end{align}

Hereafter, the thin string approximation\footnote{
This approximation corresponds to the thin wall approximation of bubble nucleation cases.} 
$\ell_s \ll \beta^{-1}$ is taken, where $\ell_s$ is the thickness of the string.
Also, the envelope approximation~\cite{Kosowsky:1991ua} is used
so that the collided strings are assumed to disappear immediately by pair annihilation.\footnote{When the strings are highly relativistic, 
the collision may lead to reformation of a pair of the strings~\cite{Myers:1991yh}.
Nevertheless, there must be a DW stretching between the reformed strings, which should make the lifetime of the pair significantly shorter and hence validate the envelope approximation.
}
Then, the energy momentum tensor (in the position space) of expanding and uncollided string loops is expressed as 
\begin{align}
    T_{ij} (x) = \rho^s (x) \hat{n} (x-x_N)_i  \hat{n} (x-x_N)_j , 
    \label{eq:em_tensor}
\end{align}
where the energy density is
\begin{align}
    \rho^s (x) = \pi r_s (t)^2 \frac{\kappa \rho_{\rm re}}{2\pi r_s (t) \ell_s} \ \ \ \  (r_s (t) < |\Vec{x} - \Vec{x}_N| < r_s (t) + \ell_s ) , \label{eq:rho-string}
\end{align}
and $\rho^s (x) = 0$ otherwise.
Here, $\rho_{\rm re}$ is the energy density released by expanding string loops. 
The efficiency factor, $\kappa$, is the fraction of kinetic energy density coming from the released energy density $\rho_{\rm re}$~\cite{Kamionkowski:1993fg}.
$x_N$ is the center of the nucleation point of the string loop, $\hat{n} (x)_i$ indicates the unit vector in the  direction of $x_i$, $\hat{n}(x)_i \equiv x_i/|\vec{x}|$, and $r_s (t)$ is the radius of the string loop. 
The initial radius is here assumed to be negligible.
Note that we do not consider the GW radiated from the remaining DWs.
When the DWs oscillate drastically,
there might be additional contribution to the energy-momentum tensor Eq.~\eqref{eq:em_tensor}.

The nucleated string loop will expand as a perfect circle until collision with other loops. 
Note that in contrast to the GW in FOPT, this expanding loops do radiate GWs even before the collision.
This can be seen from the fact that the quadrupole moment of the energy density of the single loop does not vanish.
See Appendix~\ref{app-quadrupole} for the explicit calculation.

The energy fraction of the GW is 
\begin{align}
    \Omega_{\rm GW} (t_*,k_*) \equiv \frac{1}{\rho_{\rm tot} (t_*)} \frac{d \rho_{\rm GW} (k_*)}{d \ln k_*} ,
\end{align}
where $\rho_{\rm tot} (t_*)= \rho_{\rm re} + \rho_{\rm pla} (t_*)$ is the total energy density and $\rho_{\rm pla} (t_*)$ is the plasma energy density in the background.
The gravitational wave abundance $\Omega_{\rm GW} (t_*,k_*)$ is rewritten by defining a dimensionless spectrum function $\Delta (k/\beta)$ as
\begin{align}
    \label{eq:relic2delta}
    \Omega_{\rm GW} (t_*,k_*) = & \kappa^2 \left( \frac{H_*}{\beta} \right)^2 \left( \frac{\alpha (t_*)}{1+\alpha (t_*)}  \right)^{2} \Delta (k_*/\beta) \, ,
\end{align}
    \begin{align}
    \Delta (k/\beta) = & \frac{3}{8\pi G} \frac{\beta^2 \rho_{\rm tot} (t)}{\kappa^2 \rho_{\rm re}^2} \Omega_{\rm GW} (t,k) , \\
    = & \frac{3}{4\pi^2} \frac{\beta^2 k^3}{\kappa^2 \rho_{\rm re}^2} \int_{-\infty}^{\infty} dt_1 \int_{-\infty}^{\infty} dt_2 \cos{(k(t_1 - t_2))} \Pi(t_1,t_2,k) ,
    \label{eq:Delta}
\end{align}
where $H_*^2 = 8\pi G \rho_{\rm tot} (t_*)/3$, $\alpha (t_*) = \rho_{\rm re} /\rho_{\rm pla} (t_*)$, and 
$ \Pi(t_1,t_2,k)$ can be explicitly rewritten as
\begin{align}
    \Pi(t_1,t_2,k) = & K_{ijkl} (\hat{k}) K_{ijmn} (\hat{k}) \int d^3 r e^{i \Vec{k} \cdot \Vec{r}} \langle  T_{kl} T_{mn} \rangle_{\rm ens} (t_1,t_2,r) \, .
    \label{eq:Pi_b}
\end{align} 
Here, $r=|\Vec{r}_1-\Vec{r}_2|$ is the spatial distance between two evaluation points.
We have omitted the labels of the DWs for notational simplicity.
As seen below, $\Delta$ depends on $k$ only through a form $k/\beta$.
Note that the runaway case is assumed and the initial radius of the nucleated loops is ignored.
Therefore, the speed of the expanding string loop is approximated by that of light.

The ensemble average $\langle  T_{kl} T_{mn} \rangle_{\rm ens} (t_1,t_2,r)$ has spatial dependence only through the distance $r$
due to the homogeneity and isotropy.
The value of the energy momentum tensor at a space-time point $(t_i, \Vec{r}_i)$ is determined by the time evolution of the expanding string loops that move with the speed of light.
Therefore, the calculation reduces to the geometric consideration determined by light cone structures.

\begin{figure}[t]
    \centering
    \includegraphics[width=0.7\textwidth]{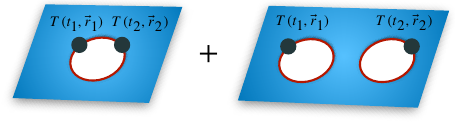}
    \caption{
    Illustration of two contributions to the two-point correlation function of the energy-momentum tensors.
    The two evaluation points can lie on one string loop (left) or on two loops (right).
    Although we take $t_1=t_2$ for the illustration in the figure,
    they can be different in general.
    Note that the GWs can be radiated even when the loops do not yet collide to the other loops.
    }
    \label{fig:single-double}
\end{figure}  

There are two contributions of energy momentum tensors depending on where the evaluation points of the energy-momentum tensors lies.
One comes from a single expanding string loop,
in which they lie on the same expanding loop,
whereas the other one comes from two expanding string loops, in which they lie on different expanding loops.
See Fig.~\ref{fig:single-double} for the illustration.
Thus we can divide the correlation function into two parts as
\begin{align}
    \langle T_{kl} T_{mn} \rangle_{\rm ens} (t_1,t_2,r)=\langle T_{kl} T_{mn} \rangle_{\rm ens}^{\rm single} (t_1,t_2,r)+\langle T_{kl} T_{mn} \rangle_{\rm ens}^{\rm double} (t_1,t_2,r) .
\end{align}
Correspondingly, $\Delta (k_*/\beta)$ is also divided as
\begin{align}
    \Delta (k/\beta) &=  \Delta^\mr{single} (k/\beta) + \Delta^\mr{double} (k/\beta) \label{eq:sum-Delta}
\end{align}
with
\begin{align}
 \Delta^\mr{single} (k/\beta) \equiv & \frac{3}{4\pi^2} \frac{\beta^2 k^3}{\kappa^2 \rho_{\rm re}^2} \int_{-\infty}^{\infty} dt_1 \int_{-\infty}^{\infty} dt_2 \cos{(k(t_1 - t_2))} \Pi(t_1,t_2,k)^{\rm single}  \label{eq:Delta-single} \\
 \Delta^\mr{double} (k/\beta) \equiv & \frac{3}{4\pi^2} \frac{\beta^2 k^3}{\kappa^2 \rho_{\rm re}^2} \int_{-\infty}^{\infty} dt_1 \int_{-\infty}^{\infty} dt_2 \cos{(k(t_1 - t_2))} \Pi(t_1,t_2,k)^\mr{double}   \, . \label{eq:Delta-double}
\end{align}
While $\langle T_{kl} T_{mn} \rangle_{\rm ens}^{\rm single}(t_1,t_2,r)$ contains the integration over one nucleation point corresponding to the single expanding loop,
$\langle T_{kl} T_{mn} \rangle_{\rm ens}^{\rm double}(t_1,t_2,r)$ contains the integration over two nucleation points as seen below.
It is shown later that the contribution from two expanding string loops dominates the gravitational wave spectrum.
Thus we first consider the case of two expanding string loops.
Although the two-loop contribution contains that of \textit{uncollided} expanding string loops,
we call this contribution a double string-loop contribution
following the terminology introduced in Ref.~\cite{Jinno:2016vai}.

\subsection{Contribution from two expanding string loops}
\label{sec:double}
\subsubsection{Ensemble average}

\begin{figure}[t]
    \centering
    \includegraphics[width=0.75\textwidth]{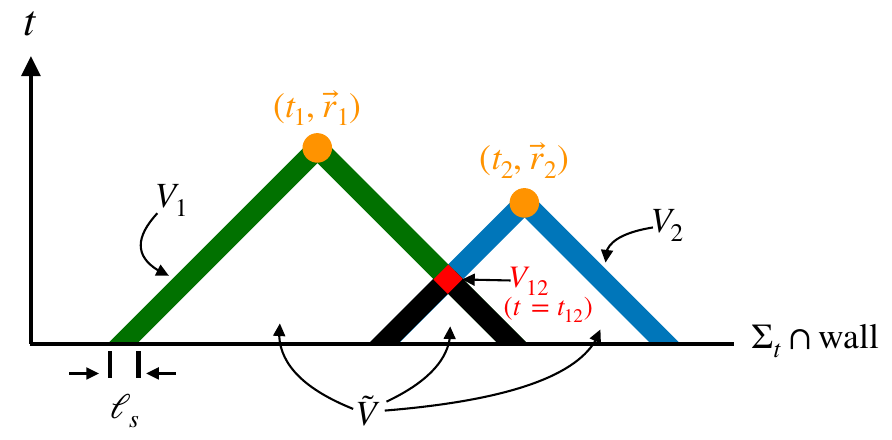} \\[4ex]
    \includegraphics[width=0.6\textwidth]{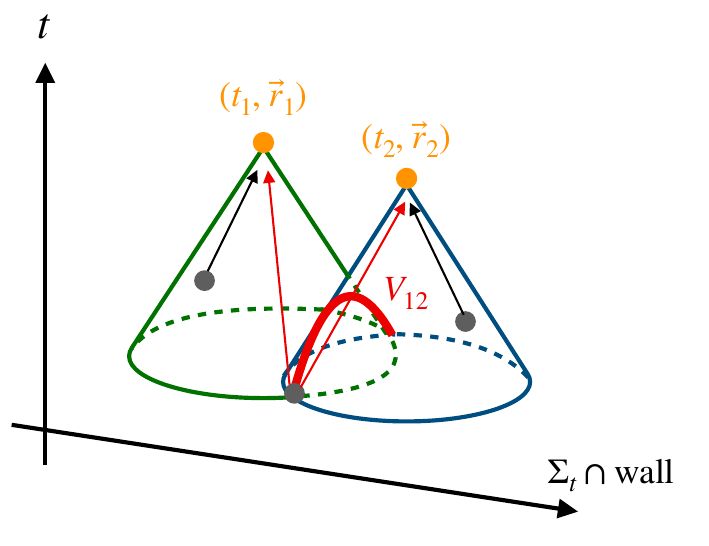}
    \caption{
    Past light cones of evaluation points $(t_i,\Vec{r}_i)$ with the width $\ell_s$. 
Top: the horizontal direction represents the two dimensional intersection of the DW and the constant-time hypersurface $\Sigma_t$.
 Nucleation points of expanding string loops must be on the surface of the past light cones with the width $\ell_s$ because of the envelope approximation.
 For a two-loop contribution, one nucleation point must be in the $2+1$ dimensional green colored region $V_1$ and the other point must be in the blue colored one $V_2$.
 For single-loop contribution, the nucleation point must be in the intersection of the two light cones $V_{12}$, shown as the red region.
 The region inside of the two light cones is denoted by $\tilde{V}$, which includes the black region.
Bottom: $2+1$ dimensional version of the top figure. 
 Gray circles indicate the nucleation points and 
 arrows represent how the expanding string loops reach the evaluation points.
 The red arch indicates $V_{12}$.
 On the time slice with $t>t_{12} \equiv (t_1+t_2-r)/2$, there is no intersection of the two light cones.}
 \label{fig:light_cone}
\end{figure}  

To proceed with our calculation, 
we need some geometric consideration.
Fig.~\ref{fig:light_cone} shows the notation of quantities for the past light cones of evaluation points $(t_1,\vec{r}_1)$ and $(t_2,\vec{r}_2)$.
As stated above, the energy-momentum tensor $T_{ij}$ lies only on the DW since the loops expand on it,
which results in that it is sufficient to consider the light cones in $2+1$ dimensional space-time ignoring the direction orthogonal to the DW.
Note that the spatial integration in Eq.~\eqref{eq:Pi_b} should be performed in the three dimensions.

The ensemble average is given by \\
(i) the value of energy momentum tensors which is on the DW, $T_{kl} (t_1,\Vec{r}_1)T_{mn} (t_2,\Vec{r}_2)$ \\ 
(ii-a) a probability that two evaluation points are on the DW, i.e., no loop is nucleated inside the two past light cones, $P (t_1,t_2,r)$ (envelope approximation)\\ 
(ii-b) a probability that each of two string loops\footnote{
In the thin string approximation, multiple string loops cannot contribute to one evaluation point.}
is nucleated in the $2+1$ dimensional green region $V_1$ 
and the blue region $V_2$ in Fig.~\ref{fig:light_cone},
$\int dt_{n1} dt_{n2} \int_{A_1} d^2 z_1 \int_{A_2} d^2 z_2 \Gamma (t_{n1}) \Gamma (t_{n2})$\\ 
(ii-c) a probability that the two evaluation points lie on the randomly distributed DW, $C_{\rm DW}(r)$. 

Here, the regions $V_1$ and $V_2$ correspond to the surfaces of the past light cones of the two evaluation points with the width $\ell_s$,
in which the black region in Fig.~\ref{fig:light_cone} is excluded by the condition (ii-a).
The integral $\int_{A_i}$ is over the area $A_i$ which represents the area of the intersection among $V_i$, the constant-time hypersurface $\Sigma_t$, and the DW.
(See Fig.~\ref{fig:plane_geo}.)
The time variable $t_{ni}$ represents the time of the loop nucleation point.

Then, the ensemble average from two expanding string loops is given by
\begin{align}
    \langle T_{kl} T_{mn} \rangle_{\rm ens}^{\rm double} (t_1,t_2,r) = C_{\rm DW}(r) P (t_1,t_2,r) & \int_{-\infty}^{t_{1}} dt_{n1} \Gamma (t_{n1}) \int_{A_1} d^2 z_1 T_{kl} (t_1,\Vec{r}) \nonumber \\ 
    \times & \int_{-\infty}^{t_{2}} dt_{n2} \Gamma (t_{n2}) \int_{A_2} d^2 z_2 T_{mn} (t_2,\Vec{r}) \, . \label{eq:TT-double}
\end{align}
The envelope approximation allows the energy to be non-zero
only when the two evaluation points are spacelike $r > |t_d|$, where $t_d = t_1 - t_2$.
Note that the two integrations about the expanding string loops 1 and 2 can be factorized into each integration as is shown in Eq.~\eqref{eq:TT-double}
because they depend on different nucleation times $t_{n1}$ and $t_{n2}$.

Comparing to the usual bubble expansion/collision, 
one should note that 
there is no spherical symmetry in the current case,
so that the uncollided loops can radiate the GW,
which corresponds to the fact that
the integration range $t_{n1(2)}>t_{12}$ gives non-zero contribution to the ensemble average. 
This dominates the GW spectrum as seen below.
Each component is calculated in the following.

\subsubsection{Random configuration of domain walls}
\label{sec:configuration_of_DW}

Let us obtain a probability that two points $\vec{r}_1,\vec{r}_2$ lie on a DW that is randomly distributed in the Hubble patch.
As the first step, let us assume that there exists only one DW within one Hubble patch.
We can rephrase this problem as the probability that a needle (or segment) that is randomly distributed in the three-dimensional space lies on a fixed DW on $y=0$.
This is similar to the 3D version of Buffon's needle.
Let the $y$ coordinate of the center of the needle be $h$ and two angles be $\theta \in [0,\pi/2)$ (angle from the DW) and $\phi\in [0,\pi)$ (angle around $y$ axes),
and $\theta\equiv \arctan (y/\sqrt{x^2+z^2})$.
The density probability functions of $h,\theta,\phi$ are given as
\begin{align}
    f_h(h)& \simeq \frac{1}{d_H} \\
    f_\theta(\theta) &= \sin \theta \\
    f_\phi(\phi) &= \frac{1}{\pi} \, ,
\end{align}
where $d_H$ is the Hubble length $d_H = 1/H_\ast$ since only one DW exists in the Hubble patch.
The two endpoints lie in the DW with its width $d_{\rm DW}$ if and only if the relation
\begin{equation}
    \left|h \pm \frac{r}{2}\cos \theta \right| \leq \frac{d_{\rm DW}}{2}
\end{equation}
is satisfied for both signs $\pm$.
Thus, we get the probability as
\begin{align}
   C_{\rm DW}^{(0)}(r) &= \int _0^\pi d \phi \int _{\cos^{-1}(d_{\rm DW}/r)}^{\pi/2} d\theta  \int _{-\frac{d_{\rm DW}}{2} + \frac{r}{2} \cos\theta}^{\frac{d_{\rm DW}}{2} - \frac{r}{2} \cos\theta} dh \,\frac{\sin\theta}{\pi d_H} \\
      &= \frac{d_{\rm DW}^2}{2 r d_H} .
\end{align}
Then, taking into account the number of the multiple DWs $N_\mr{walls}$ in the Hubble patch,
the full probability $C_{\rm DW}(r)$ is expressed as 
\begin{align}
    C_{\rm DW}(r) = & N_{\rm walls} C_{\rm DW}^{(0)}(r) , \\
    = & N_{\rm walls} \frac{d_{\rm DW}^2}{{2 r d_H}}.
\end{align}
The value of $N_{\rm walls}$ depends on cosmological scenarios
which will be argued in Sec.~\ref{sec:present-spectrum}.

\subsubsection{Probability of the domain wall to remain}

\begin{figure}[t]
    \centering
    \includegraphics[width=0.7\textwidth]{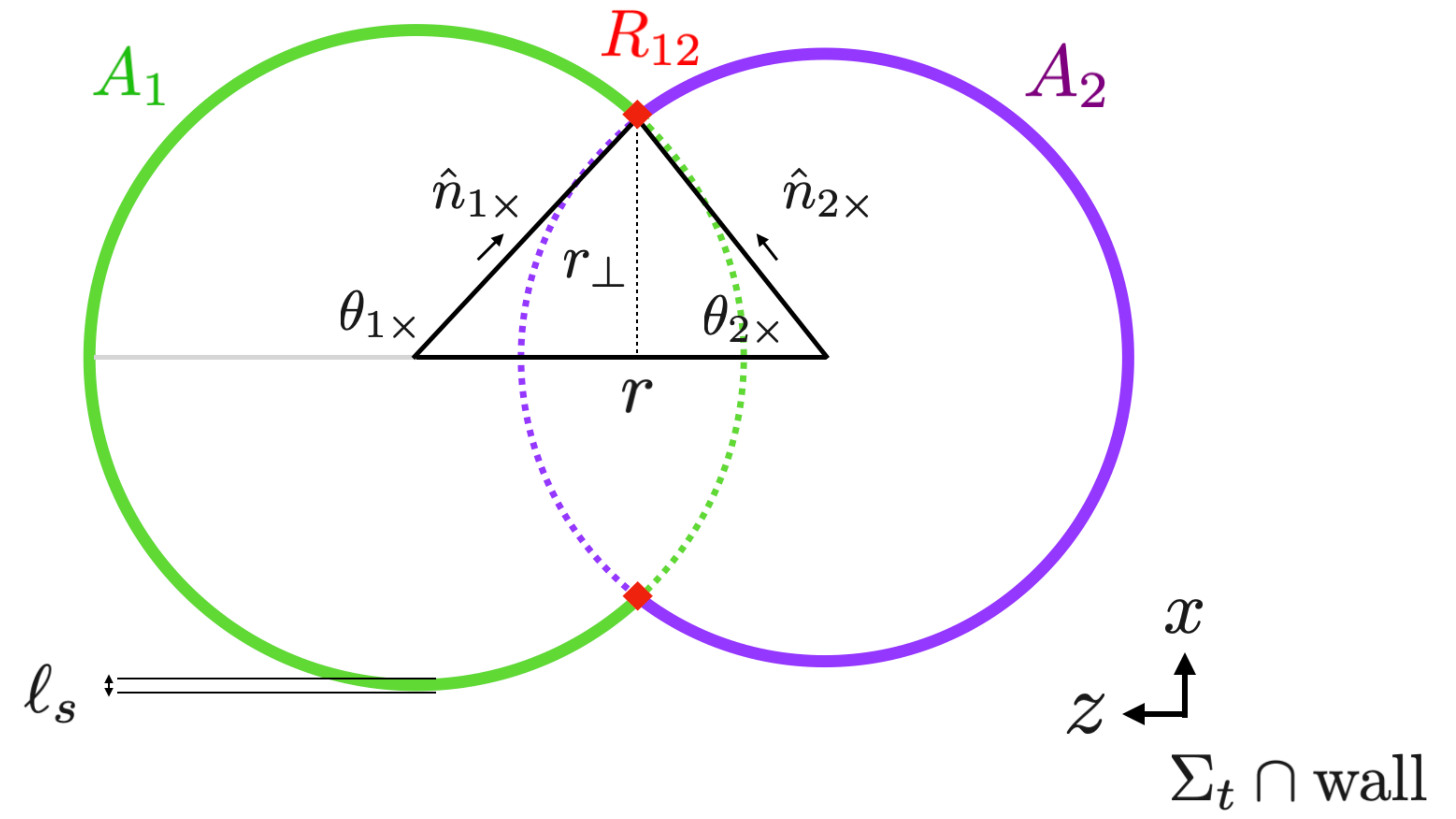}
    \caption{
    Figure of geometry for calculation on the constant-time hypersurface $\Sigma_t$.
    This is equivalent to a constant-time slice of Fig.~\ref{fig:light_cone}.
    The green left and purple right circles are associated with the past light cones of $(t_1,\vec{r}_1)$ and $(t_2,\vec{r}_2)$, respectively.
    }
    \label{fig:plane_geo}
\end{figure}  

The probability $P (t_1,t_2,r)$ is obtained by calculating the $2+1$ dimensional volume $\tilde{V}$ of the union of the inside of the two past light cones for the evaluation points shown in Fig.~\ref{fig:light_cone}.
The volume $\tilde{V}$ is obtained by integrating with respect to time
the area of the union of the inside of the two circles shown in Fig.~\ref{fig:plane_geo}.

Dividing $\tilde{V}$ into infinitesimal $2+1$ dimensional regions labeled by $i$, $\tilde{V}^i$,
the probability that no loop is nucleated in $\tilde{V}$ is expressed as
\begin{align}
    P (t_1,t_2,r) = & \prod_i (1- \Gamma (t) d \tilde{V}^i) , \\
    = & e^{-I (x_1,x_2)} \, ,
\end{align}
with 
\begin{align}
 I(x_1,x_2) \equiv \int _{\tilde{V}} d^3z \, \Gamma(t) \, .
\end{align}

To proceed with the calculation, one should note that 
the two circles on $\Sigma_t$ can be either overlapped or not, depending on $t$.
(See Figs.~\ref{fig:light_cone} and \ref{fig:plane_geo}.)
Thus we divide the integration region with respect to $t$ as follows:
\begin{align}
\begin{cases}
    \text{overlapped regime: } & -\infty<t<t_{12}\\
    \text{non-overlapped regime: } &t_{12}<t<t_{1(2)}
    \end{cases} \label{eq:overlap-unoverlap}
\end{align}
where the time $t_{12} = (t_1+t_2-r)/2$ represents when the two past light cones meet.
The integration over the non-overlapped regime,
which has a shape of two cones,
can be performed analytically.
Then, the above expression is calculated as
\begin{align}
    I (x_1,x_2) = & \Gamma (T) \mathcal{I} (r, t_d) , \\
    \mathcal{I} (r, t_d) = & \frac{1}{2 \beta^3} \left[ 8 \pi \cosh{\left(\frac{\beta t_d}{2}\right)} + \beta^2 r \sqrt{r^2 - t_d^2} K_1 \left( \frac{\beta r}{2} \right) \right] \nonumber \\
    & - \frac{1}{4} \int_{-\infty}^{-r/2} dt_{T} \, e^{\beta t_T} \left[ (2 t_{T} - t_d)^2   \left( \pi - \cos^{-1} \left( \frac{r^2-2t_d t_T}{r(2t_T - t_d)} \right) \right) \right. \nonumber \\
    & \hspace{9em} \left. + (2 t_{T} + t_d)^2 \left( \pi -  \cos^{-1} \left( \frac{r^2 + 2t_d t_T}{r(2t_T + t_d)} \right) \right) \right] ,
    \label{eq:false_vac_prob_factor}
\end{align}
where $K_1 (x)$ is the modified Bessel function of second kind.
For the overlapped regime $-\infty <t<t_{12}$,
we have shifted the time variable as $t = t_T + T$, leading to the integration range $-\infty<t_T < -r/2$ with $T \equiv (t_1 + t_2)/2$ in the second and third lines.
The above expression of $\mathcal{I} (r, t_d)$ is invariant under $t_d \to -t_d$.
The integration in Eq.~\eqref{eq:false_vac_prob_factor} is well fitted by $e^{-r/2-t_d}$ multiplied with the polynomials of $r$ and $t_d$ up to fifth order.

\subsubsection{Evaluation of $\Delta^\mr{double}$}
We assume the DW plane to be the $xz$ plane in the three-dimensional coordinate space.
Then, the vector $\Vec{r}$ is on the $xz$ plane and the direction of $\Vec{r} = \Vec{r}_1-\Vec{r}_2$ can be taken as the $z$ axis without loss of generality.

The areas on which the string loops 1(2) can be nucleated are (see Fig.~\ref{fig:plane_geo}.)
\begin{align}
    \int_{A_{1}} d^2 z_1 = & \ell_s
    \begin{cases}
        \int_{-\theta_{1\times}}^{\theta_{1\times}} r_{1\times} (t_{n1}) d\theta_{1} & \ \ \ ({\rm for} -\infty <t_{n1}<t_{12}) \\
        2\pi r_{1\times} (t_{n1}) & \ \ \ ({\rm for} \ \ \ t_{12}<t_{n1}<t_1)
    \end{cases} 
    , \\
    \int_{A_{2}} d^2 z_2 = & \ell_s
    \begin{cases}
        \int_{ \theta_{2\times}}^{2\pi-\theta_{2\times}} r_{2\times} (t_{n2}) d\theta_{2} & ({\rm for} -\infty <t_{n2}<t_{12}) \\
        2\pi r_{2\times} (t_{n2}) & ({\rm for} \ \ \ t_{12}<t_{n2}<t_2)
    \end{cases} 
    .
\end{align}
Here, $\times$ indicates the overlapped points of past light cones shown in Fig.~\ref{fig:plane_geo}.
$r_{1\times } (t_{n}) = t_{1} - t_{n}$ and $r_{2\times } (t_{n}) = t_{2} - t_{n}$ are the radius of the green left and purple right circles in Fig.~\ref{fig:plane_geo} at $t=t_{n}$, respectively.
The angles $\theta_{1\times}$ and $\theta_{2 \times}$ are rewritten as
\begin{align}
 \cos \theta_{1\times}=-\frac{r^2+r_{1\times}(t_{n1})^2-r_{2\times}(t_{n1})^2}{2r_{1\times}(t_{n1})r}
\end{align}
\begin{align}
 \cos \theta_{2\times}=\frac{r^2+r_{2\times}(t_{n2})^2-r_{1\times}(t_{n2})^2}{2r_{2\times}(t_{n2})r} \, .
\end{align}
(See Fig.~\ref{fig:plane_geo}.)
Note that $-\infty<t_{n1(2)} < t_{12}$ corresponds to the region in which two past light cones have intersection while $t_{12} < t_{n1(2)}<t_{1(2)}$ corresponds to non-overlapped regime,
as introduced in Eq.~\eqref{eq:overlap-unoverlap}.

For the overlapped regime, 
using Eq.~\eqref{eq:em_tensor},
the integration of the energy-momentum tensors in Eq.~\eqref{eq:TT-double} are written as 
\begin{align}
    \int_{A_1} d^2 z_1 \int_{A_2} d^2 z_2 T_{kl} (t_1,\Vec{r})T_{mn} (t_2,\Vec{r}) = & \left( \frac{\kappa \rho_{\rm re}}{2} \right)^2 \int_{-\theta_{1\times}}^{\theta_{1\times}} d\theta_{1} r_{1\times}^2 (t_{n1}) \hat{n}_{1,k}(t_{n1}) \hat{n}_{1,l}(t_{n1}) \nonumber \\
    & \ \  \times \int_{ \theta_{2\times}}^{2\pi-\theta_{2\times}} d\theta_{2} r_{2\times}^2 (t_{n2}) \hat{n}_{2,m}(t_{n2}) \hat{n}_{2,n}(t_{n2}) \, . \label{eq:EMtensor-integral}
\end{align}
Here, $\hat{n}_{1(2),k}$ is the $k$-th spatial component of unit vectors
$\hat{n}_{1(2)} = (\sin \theta_{1(2)} ,0, \cos \theta_{1(2)})$.
The above expression does not depend on $\ell_s$.
For the non-overlapped regime, $t_{12}<t_{n1(2)}<t_{1(2)}$, 
the corresponding expression is obtained by taking $\theta_{1\times}=\pi$ ($\theta_{2\times}=0$) in the above expression,
which means that the integration ranges become ``full circles''
and hence manifestly give $O(2)$ symmetric contributions.

Now let us calculate Eq.~\eqref{eq:Pi_b}.
Without loss of generality, the wave vector $\Vec{k}$ is parametrized as $\Vec{k} = k (\sin{\theta_{kr}} \cos{\phi_{kr}}, \sin{\theta_{kr}} \sin{\phi_{kr}}, \cos{\theta_{kr}})$
with angles $\theta_{kr}\in[0,\pi]$ and $\phi_{kr}\in[0,2\pi)$.
Using these definitions, 
the integral $\int d^3 r$ in Eq.~(\ref{eq:Pi_b}) is written as $\int d^3 r =\int_0^\infty r^2 dr \int_{-1}^1 d c_{kr}\int_0^{2\pi} d \phi_{kr}$, where $c_{rk} = \hat{k} \cdot \hat{r} =  \cos{\theta_{kr}}$.

Noting that the variables $t_{1(2)}$ can be explicitly written by $T=(t_1+t_2)/2$ and $t_d=t_1-t_2$ as $t_{1/2}=T\pm t_d/2$,
the TT projection in Eq.~\eqref{eq:Pi_b} is performed as
\begin{align}
  &  \int_0^{2\pi} d\phi_{kr} K_{ijkl}(\hat{k}) K_{ijmn}(\hat{k}) \langle T_{kl} T_{mn} \rangle_{\rm ens}^{\rm double}(t_1,t_2,r) \nonumber \\
= & \frac{\pi}{4} \kappa^2 \rho_{\rm re}^2 C_{\rm DW} P (t_1,t_2,r) \Gamma (T)^2 \int_{-\infty}^{t_d/2} dt_{1T} \int_{-\infty}^{-t_d/2} dt_{2T} \, r_{1\times}^2 r_{2\times}^2 \nonumber \\
    & \times e^{\beta (t_{1T} + t_{2T})} \left[ \frac{1}{2} G_0 + \frac{1}{4}(1-c_{rk}^2) G_1 + \frac{1}{16}(1-c_{rk}^2)^2 G_2 \right] ,\label{eq:TTproj-double}
\end{align}
where $t_{iT} \equiv t_{ni} - T$, and 
\begin{align}
    G_0(\theta_{1\times},\theta_{2\times}) & =  2 (\theta_{1\times} -\sin\theta_{1\times} \cos \theta_{1\times}) (\pi - \theta_{2\times} +\sin\theta_{2\times} \cos \theta_{2\times})   , \\
    G_1(\theta_{1\times},\theta_{2\times}) & = -2 G_0(\theta_{1\times},\theta_{2\times}) , \\
    G_2(\theta_{1\times},\theta_{2\times}) & = 6\theta_{1\times}(\pi-\theta_{2\times})-10\theta_{1\times} \sin\theta_{2\times}\cos\theta_{2\times} +10(\pi-\theta_{2\times})\sin\theta_{1\times}\cos\theta_{1\times} \nonumber \\
    & \hspace{2em} -38 \sin\theta_{1\times}\cos\theta_{1\times}\sin\theta_{2\times}\cos\theta_{2\times}  .
\end{align}
The functions, $G_i(\theta_{1\times},\theta_{2\times})$, have a symmetry under the exchange between $\theta_{1\times}$ and $\bar{\theta}_{2\times} = \pi-\theta_{2\times}$.

Then, substituting Eqs.~\eqref{eq:TTproj-double} and~\eqref{eq:Pi_b}
into Eq.~\eqref{eq:Delta-double},
we get the analytic formula of the spectrum function $\Delta^{\rm double} (k/\beta)$ as
\begin{align}
    \Delta^{\rm double}(k/\beta) = & \frac{3}{64\pi}\beta k^3 \int_{0}^\infty dt_d \cos (kt_d) \int_{t_d}^\infty dr \frac{r^2 C_{\rm DW}(r)}{\mathcal{I}(r,t_d)^2} \nonumber \\
    & \times \int_{-\infty}^{t_d/2} dt_{1T} \int_{-\infty}^{-t_d/2} dt_{2T} \, e^{\beta (t_{1T}+t_{2T})} \left[ \left( j_0 (kr) -2 \frac{j_1(kr)}{kr} \right) G_0' + \frac{j_2(kr)}{(kr)^2} G_2'\right] ,
\end{align}
where, $j_0(x)=\frac{1}{2}\int_{-1}^1 dc\, e^{icx}$, $j_1(x)/x=\frac{1}{4}\int_{-1}^1 dc (1-c^2)e^{icx}$, and $j_2(x)/x^2=\frac{1}{16}\int_{-1}^1 dc (1-c^2)^2e^{icx}$ are the spherical Bessel functions,
\begin{align}
    G_0'(r,t_d,t_{1T},t_{2T}) = &  \left[g_{a}' (r,t_d,t_{1T}) + g_{b}' (r,t_d,t_{1T})\right]\left[g_{a}' (r,-t_d,t_{2T}) + g_{b}' (r,-t_d,t_{2T})\right] , \\
    G_2'(r,t_d,t_{1T},t_{2T}) = & 3g_{a}'(r,t_d,t_{1T})g_{a}'(r,-t_d,t_{2T})+19g_{b}' (r,t_d,t_{1T}) g_{b}'(r,-t_d,t_{2T}) \nonumber \\
    & -5g_{a}'(r,t_d,t_{1T})g_{b}'(r,-t_d,t_{2T})-5g_{b}'(r,t_d,t_{1T})g_{a}'(r,-t_d,t_{2T})  ,
\end{align}
and
\begin{align}
    g_{a{\rm }}' (r,t_d,t_{1T}) = & 
    \begin{cases}
        (-2t_{1T}+t_d)^2  \cos^{-1} \left( \frac{r^2-2t_{1T}t_d}{r(2t_{1T}-t_d)} \right) & \ \ \ (-\infty<t_{1T} < -r/2) \\
        \pi (-2t_{1T}+t_d)^2 & \ \ \ (-r/2 < t_{1T}<t_d/2)
    \end{cases} 
    , \\
    g_{b{\rm }}' (r,t_d,t_{1T}) = & 
    \begin{cases}
        \frac{r^2-2t_{1T}t_d}{r^2}\sqrt{(4t_{1T}^2-r^2)(r^2-t_d^2)} & \ \ \ \ (-\infty<t_{1T} < -r/2) \\
        0 & \ \ \ \ (-r/2 < t_{1T}<t_d/2)
    \end{cases} 
    .
\end{align}
Here, we have used a relation $r>|t_d|$ and a symmetry of $G_i'$ under $t_d\to -t_d$.
We also changed the integration variables $t_1$ and $t_2$ into $t_d$ and $T$,
and explicitly performed the integration over $T$ by using a formula $\int_{-\infty}^\infty dy\,  z^{n} e^{-xz} = (n-1)!/x^n$ with $z=e^y$.
Note that $-\infty<t_{1(2)T} < -r/2$ corresponds to the overlapped regime while $-r/2 < t_{1T}<t_d/2$
and $-r/2 < t_{2T}<- t_d/2$ correspond to the non-overlapped regime.

The above expression of $\Delta^\mr{double}(k/\beta)$ can be further simplified by introducing
\begin{align}
    g_{a,{\rm ov}}'' (r,t_d) \equiv & \int_{-\infty}^{-r/2}dt_{1T} \, e^{\beta t_{1T}} g_{a{\rm }}' (r,t_d,t_{1T}) , \\
    g_{a, {\rm non}}'' (r,t_d) \equiv & \int_{-r/2}^{t_d/2} dt_{1T} \,e^{\beta t_{1T}} g_{a {\rm }}' (r,t_d,t_{1T}) \\ 
    = & \frac{\pi}{\beta^3} e^{-\beta r/2} \left[ 8(-1+e^{\beta (r+t_d)/2})-\beta (r+t_d)\left\{ 4+\beta (r+t_d)\right\} \right] , \\
    g_{b,{\rm ov}}'' (r,t_d) \equiv & \int_{-\infty}^{-r/2}dt_{1T} \,e^{\beta t_{1T}} g_{b{\rm }}' (r,t_d,t_{1T}) \\
    = & \frac{\sqrt{r^2-t_d^2}}{\beta} \left[ r K_1 \left( \frac{\beta r}{2} \right)+t_d K_2 \left( \frac{\beta r}{2} \right) \right]\, ,
\end{align}
where the subscripts ``$\mathrm{ov}$'' and ``$\mathrm{non}$'' mean the integral region in which the two past light cones are overlapped and non-overlapped, respectively.
Then, the spectrum function $\Delta^{\rm double}(k/\beta)$ becomes
\begin{align}
    \Delta^{\rm double}(k/\beta) = & \frac{3}{64\pi}\beta k^3 \int_{0}^\infty dt_d \cos (kt_d) \int_{t_d}^\infty dr \frac{r^2 C_{\rm DW}(r)}{\mathcal{I}(r,t_d)^2} \nonumber \\
    & \times \left[ \left( j_0 (kr) -2 \frac{j_1(kr)}{kr} + 3\frac{j_2(kr)}{(kr)^2}\right) \right. \nonumber \\
    & \ \ \times \left( g_{a,{\rm ov}}''(r,t_d)+g_{a,{\rm non}}''(r,t_d)\right)\left(g_{a,{\rm ov}}''(r,-t_d) +g_{a,{\rm non}}''(r,-t_d)\right) \nonumber \\
    & \ \ + \left( j_0 (kr) -2 \frac{j_1(kr)}{kr} -5\frac{j_2(kr)}{(kr)^2}\right) \nonumber \\
    & \ \ \times \left[ \left( g_{a,{\rm ov}}''(r,t_d)+g_{a,{\rm non}}''(r,t_d)\right) g_{b,{\rm ov}}''(r,-t_d)+ g_{b,{\rm ov}}''(r,t_d)\left(g_{a,{\rm ov}}''(r,-t_d) +g_{a,{\rm non}}''(r,-t_d)\right) \right] \nonumber \\
    & \left. + \left( j_0 (kr) -2 \frac{j_1(kr)}{kr} +19\frac{j_2(kr)}{(kr)^2}\right) g_{b,{\rm ov}}''(r,t_d)g_{b,{\rm ov}}''(r,-t_d) \right] .
    \label{eq:delta_double}
\end{align}
This is the main result of the analytic formula for $\Delta^\mathrm{double}(k/\beta)$.

The remaining integrals are performed numerically.
However, there are simply oscillating terms at large $r$ whose integration does not converge as shown in the following.
The product of the contributions from the non-overlapped regime for both string loops, $g_{a,{\rm non}}''(r,t_d) g_{a,{\rm non}}''(r,-t_d)$,
is separated as
\begin{align}
    g_{a,{\rm non}}''(r,t_d) g_{a,{\rm non}}''(r,-t_d) = & g_{\rm IR}(r,t_d) + g_{{\rm conv}}(r,t_d) , \label{eq:non-overlap}
\end{align}
where
\begin{align}
    g_{{\rm IR}}(r,t_d) = & 64\pi^2   , \label{eq:gIR}\\
    g_{{\rm conv}}(r,t_d) = & \pi^2 e^{-r} \left[ (8+4r+r^2)^2-2r(4+r)t_d^2 +t_d^4 \right] \nonumber \\
    & -16\pi^2 e^{-\frac{r}{2}} \left( (8+4r+r^2+t_d^2) \cosh{\frac{t_d}{2}} -2(2+r)t_d \sinh{\frac{t_d}{2}}  \right)    .
\end{align}
Note that $g_{{\rm conv}}(r,t_d)$ exponentially decays at $r\to \infty$
whereas $g_{{\rm IR}}(r,t_d)$ is a constant and not well-behaved.
Thus, let us take a closer look at the contribution from $g_{{\rm IR}}(r,t_d)$, which is further separated as
\begin{align}
\left. \Delta^\mathrm{double}(k/\beta) \right|_{g_{{\rm IR}}\text{ part}} =  \Delta^{\rm double}_{{\rm IR0}}(k/\beta) + \Delta^{\rm double}_{{\rm IR12}}(k/\beta)
\end{align}
with
\begin{align}
    \Delta^{\rm double}_{{\rm IR0}}(k/\beta) = & {3\pi}{}\beta k^3 \int_{0}^\infty dt_d \cos (kt_d) \int_{t_d}^\infty dr \frac{r^2 C_{\rm DW}(r)}{\mathcal{I}(r,t_d)^2}   j_0 (kr) ,\label{eq:ddc0} \\
    \Delta^{\rm double}_{{\rm IR12}}(k/\beta) = & {3\pi}{}\beta k^3 \int_{0}^\infty dt_d \cos (kt_d) \int_{t_d}^\infty dr \frac{r^2 C_{\rm DW}(r)}{\mathcal{I}(r,t_d)^2} \left[  -2 \frac{j_1(kr)}{kr} + 3\frac{j_2(kr)}{(kr)^2} \right]  . 
    \label{eq:ddc12}
\end{align}

One can see that the second term $\Delta^{\rm double}_{{\rm IR12}}(k/\beta)$ is a convergent integral due to decaying polynomials of $r$
 while the first term $\Delta^{\rm double}_{{\rm IR0}}(k/\beta)$ contains simply oscillating terms at large $r$.
Indeed, at $r\to \infty$, its integrand is approximated as
\begin{align}
    \Delta^{\rm double}_{{\rm IR0}}(k/\beta) \overset{r\to\infty}{\to} & ~{3\pi}{}\beta k^3 \int_{0}^\infty dt_d \cos (kt_d) \int_{t_d}^\infty dr r^2 C_{\rm DW}(r)  \frac{1}{\mathcal{I}(\infty,t_d)^2}   j_0 (kr) , \\
    = & \frac{3N_{\rm walls}d_{\rm DW}^2 H_* k^2}{32\pi}  \int_{0}^\infty dr_{t_d} \left[\left( 1+\frac{2\pi k/\beta}{\sinh\frac{2\pi k}{\beta}} \right) \sin (kr_{t_d}) + \beta^{-1}k \chi(k/\beta)\cos (kr) \right] , 
\end{align}
with $r_{t_d} \equiv r - t_d$ and 
\begin{align}
    \chi(k/\beta) = & H_{ik/\beta}+H_{-ik/\beta}-H_{-\frac{1}{2}+ik/\beta} -H_{-\frac{1}{2}-ik/\beta} , \\
    H_y = & \int_0^1 dx \frac{1-x^y}{1-x} \, ,
\end{align}
where $H_y$ is the harmonic number of integral representation for the complex plane.
Introducing an IR cutoff $L$ for the integration, it gives
\begin{align}
    \Delta^{\rm double}_{{\rm IR0}}(k/\beta) \overset{r\to\infty}{\to} &~ \frac{3N_{\rm walls}d_{\rm DW}^2 H_*k}{32\pi} \left[\left( 1+\frac{2\pi k/\beta}{\sinh\frac{2\pi k}{\beta}} \right) (1- \cos (kL)) + \beta^{-1}k \chi(k/\beta)\sin (kL) \right]  \, ,
\end{align}
which is not convergent but depends on the IR cutoff $L$ explicitly.

Therefore, we rewrite the divergent integral $\Delta^{\rm double}_{{\rm IR0}}(k/\beta)$, Eq.~\eqref{eq:ddc0}, using the following trick:
\begin{align}
\Delta^{\rm double}_{{\rm IR0}}(k/\beta) 
= & {3\pi}{}\beta k^3 \int_{0}^\infty dt_d \cos (kt_d) \int_{t_d}^\infty dr r^2 C_{\rm DW}(r) \left( \frac{1}{\mathcal{I}(r,t_d)^2} - \frac{1}{\mathcal{I}(\infty,t_d)^2} +\frac{1}{\mathcal{I}(\infty,t_d)^2}\right)  j_0 (kr)  \\
= & {3\pi}{}\beta k^3 \int_{0}^\infty dt_d \cos (kt_d) \int_{t_d}^\infty dr r^2 C_{\rm DW}(r) \left( \frac{1}{\mathcal{I}(r,t_d)^2} - \frac{1}{\mathcal{I}(\infty,t_d)^2} \right)  j_0 (kr) \nonumber \\ 
& + \frac{3N_{\rm walls}d_{\rm DW}^2 H_*k}{32\pi} \left[\left( 1+\frac{2\pi k/\beta}{\sinh\frac{2\pi k}{\beta}} \right) (1- \cos (kL)) + \beta^{-1}k \chi(k/\beta)\sin (kL) \right]  \, . \label{eq:subt-ddc0}
\end{align}
Due to the subtraction within the parentheses in the first line in Eq.~\eqref{eq:subt-ddc0},
the numerical integration with respect to $r$ is well convergent and easy to evaluate.
On the other hand, the second line explicitly depends on the IR cutoff $L$.
Note that the IR cutoff $L$ corresponds to the curvature radius of the DW,
beyond which our approximation of the DW to be planar is not reliable.
Therefore its appearance does not necessarily mean an inconsistency of the calculation.

We here take the average over $L$ so that the final result does not depend on $L$.
Thus, we get
\begin{align}
\Delta^{\rm double}_{{\rm IR0}}(k/\beta) 
= & {3\pi}{}\beta k^3 \int_{0}^\infty dt_d \cos (kt_d) \int_{t_d}^\infty dr r^2 C_{\rm DW}(r) \left( \frac{1}{\mathcal{I}(r,t_d)^2} - \frac{1}{\mathcal{I}(\infty,t_d)^2} \right)  j_0 (kr) \nonumber \\ 
& + \frac{3N_{\rm walls}d_{\rm DW}^2 H_*k}{32\pi} \left( 1+\frac{2\pi k/\beta}{\sinh\frac{2\pi k}{\beta}} \right) .
\label{eq:ddc0n}
\end{align}
One may consider that this prescription of $L$ looks somewhat ad hoc.
Nevertheless, it is justified by the causality argument.
If a different choice of $L$ was taken,
$\Delta^\mathrm{double}(k/\beta)$ and hence the final result of $\Delta(k/\beta)$ would be proportional to $k^1$ in the IR regime (see Fig.~\ref{fig:delta_double_const} in Appendix~\ref{sec:k_cancel}.),
which is inconsistent with the requirement from the causality~\cite{Caprini:2009fx}.
Due to averaging over $L$, 
the $k^1$ dependence is canceled with the other contributions and hence 
the final result recovers the $k^3$ dependence of the spectrum at a small $k$ region.

Note that the second line in Eq.~\eqref{eq:ddc0n} gives a linear contribution $k^1$ in the UV regime of $k$.
An essential reason for this linear dependence is that this contribution comes from the constant-in-time source, $g_\mr{IR}(r,t_d)$~\eqref{eq:gIR}.
In this sense, this behavior is consistent with the studies on the linear dependence in Ref.~\cite{RoperPol:2022iel}
although the technical details are different.
Note also that 
this constant-in-time source $g_\mr{IR}(r,t_d)$
originates from the non-overlapped regime 
Eq.~\eqref{eq:non-overlap},
which corresponds to the $O(2)$-symmetric contribution as stated below Eq.~\eqref{eq:EMtensor-integral}.
A counterpart of this contribution in cases of bubble collision/expansion of usual FOPTs
is a $O(3)$-symmetric contribution and does vanish obviously.
Therefore this linear behavior is peculiar to our case.

\subsection{Contribution from single expanding string loop}
\label{sec:single}

Next, we consider the contribution from single string loop.
The ensemble average is given by \\
(i) the value of the energy momentum tensors on the DW, $T_{kl} (t_1,\Vec{r}_1)T_{mn} (t_2,\Vec{r}_2)$ \\
(ii-a) the probability that two evaluation points are on the DW, i.e., no loop is nucleated inside the two past light cones shown in Fig.~\ref{fig:light_cone} (envelope approximation), $P (t_1,t_2,r)$\\ 
(ii-b$'$) a probability that one string loop is nucleated in the red arch region $V_{12}$ in the bottom figure of Fig.~\ref{fig:light_cone}, $\int dt_n \int_{R_{12}} d^2 z \Gamma (t_n)$ \\ 
(ii-c) the probability that the two evaluation points lie on the randomly distributed DW, $C_{\rm DW}(r)$.\\
Then, let us consider the ensemble average for a single expanding string loop:
\begin{align}
    \langle T_{kl} T_{mn} \rangle_{\rm ens}^{\rm single} (t_1,t_2,r) = & C_{\rm DW}(r) P (t_1,t_2,r) \int_{-\infty}^{t_{12}} dt_n \Gamma (t_n) \int_{R_{12}} d^2 z T_{kl} (t_1,\Vec{r})T_{mn} (t_2,\Vec{r}) \, .
\end{align}
The region $R_{12}$ is the intersection among $V_{12}$, $\Sigma_t$, and the DW,
and consists of two separated diamond red regions in Fig.~\ref{fig:plane_geo}. 
The area of $R_{12}$ is given as $\int_{R_{12}} d^2 z\approx 2\ell_s^2/\sin (\theta_{1\times} - \theta_{2\times}) = 2\ell_s^2 r_{1\times}(t_n) r_{2\times}(t_n)/(r_{\perp} r)$, where $r_{\perp} =r_{1\times} \sin \theta_{1\times}=r_{2\times} \sin \theta_{2\times}$ is the distance between $R_{12}$ and $\Vec{r}$.
For the single loop case, only one nucleation time appears in the calculation, instead of two.

In contrast to the two-string contribution in Sec.~\ref{sec:double}, 
the integration over the nucleation time $t_n$ does not contain the contribution from the non-overlapped regime but has the upper limit $t_{12}$
because the two light cones do not have any intersection for $t>t_{12}$.
This is a crucial reason why the single loop case does not give the dominant contribution in the GW spectrum, 
as seen below.

Using Eq.~\eqref{eq:em_tensor},
we obtain
\begin{align}
    T_{kl} (t_1,\Vec{r})T_{mn} (t_2,\Vec{r}) = & \left( \frac{\kappa \rho_{\rm re}}{2 \ell_s} \right)^2 r_{1\times} (t_n) r_{2\times} (t_n) (N_\times)_{kl,mn} , 
\end{align}
with
\begin{align}
 (N_\times)_{kl,mn} \equiv \hat{n}_{1\times ,k} \, \hat{n}_{1\times ,l} \, \hat{n}_{2\times,m} \, \hat{n}_{2\times,n} \, ,
\end{align}
where $\hat{n}_{1\times}$ and $\hat{n}_{2\times}$ are unit vectors pointing from the center of each circle to the diamond region $R_{12}$ in Fig.~\ref{fig:plane_geo};
$\hat{n}_{1(2)\times} = (\sin \theta_{1(2)\times} ,0, \cos \theta_{1(2)\times})$.
Performing the TT projection leads to
\begin{align}
    K_{ijkl} (\hat{k}) K_{ijmn} (\hat{k}) \langle T_{kl} T_{mn} \rangle_{\rm ens}^{\rm single} (t_1,t_2,r) = & \frac{\kappa^2 \rho_{\rm re}^2}{2}  C_{\rm DW}(r) P (t_1,t_2,r) \int_{-\infty}^{t_{12}} dt_n \Gamma (t_n) \nonumber \\
    & \times \frac{r_{1\times} (t_n)^2 r_{2\times} (t_n)^2}{r \, r_{\perp}} K_{klmn}(\hat{k}) (N_\times)_{kl,mn}\, .
    \label{eq:KKTT_sub}
\end{align}
This expression does not depend on $\ell_s$.

By using the definition of $\Vec{k} = k (\sin{\theta_{kr}} \cos{\phi_{kr}}, \sin{\theta_{kr}} \sin{\phi_{kr}}, \cos{\theta_{kr}})$,
the above expression can be integrated to be
\begin{align}
    \int_0^{2\pi} d\phi_{kr} K_{ijkl}  K_{ijmn}  \langle T_{kl} T_{mn} \rangle_{\rm ens}^{\rm single} = & \frac{\pi}{2} \kappa^2 \rho_{\rm re}^2 C_{\rm DW}(r) P (t_1,t_2,r) \Gamma (T) \int_{-\infty}^{-r/2} dt_T \frac{r_{1\times}^2 r_{2\times}^2 }{r^5 r_{\perp}}  \nonumber \\
    & \times e^{\beta t_T} \left[ \frac{1}{2} F_0 + \frac{1}{4}(1-c_{rk}^2) F_1 + \frac{1}{16}(1-c_{rk}^2)^2 F_2 \right] ,\label{eq:TTproj-single}
\end{align}
where $t_T = t_n - T$ and 
\begin{align}
    F_0(\theta_{1\times},\theta_{2\times}) = & 2 \sin^2 \theta_{1\times} \sin^2 \theta_{2\times}   , \\
    F_1(\theta_{1\times},\theta_{2\times}) = & -4 \sin^2 \theta_{1\times} \sin^2 \theta_{2\times} + 16 \sin \theta_{1\times} \cos \theta_{1\times} \sin \theta_{2\times} \cos \theta_{2\times} , \\
    F_2(\theta_{1\times},\theta_{2\times}) = & \frac{1}{2} [3+5\cos (2\theta_{1\times}) +5\cos (2\theta_{2\times}) \nonumber \\
    &\ \ +19 \cos (2\theta_{1\times}) \cos (2\theta_{2\times})-16 \sin (2\theta_{1\times}) \sin (2\theta_{2\times})] .
\end{align}

Then,
substituting Eqs.~\eqref{eq:TTproj-single} and~\eqref{eq:Pi_b}
into Eq.~\eqref{eq:Delta-single},
the spectrum function $\Delta^{\rm single}$ is calculated as
\begin{align}
    \Delta^{\rm single}(k/\beta) =  \frac{3}{4\pi} \beta k^3 \int_{0}^{\infty} dt_d \cos (k t_d) \int_{t_d}^{\infty} dr \frac{C_{\rm DW}(r)}{r^3 \mathcal{I}(r,t_d)} \left[ j_0 (kr) F_0' +  \frac{j_1(kr)}{kr} F_1' + \frac{j_2(kr)}{(kr)^2} F_2' \right] ,\label{eq:Delta-single-result}
\end{align}
where
we have used $r_{\perp} = \sqrt{(4t_T^2-r^2)(r^2-t_d^2)}/(2r)$, 
an invariant property of $\Delta^{\rm single}(k/\beta)$ under $t_d \to -t_d$, 
and the relation $r>|t_d|$.
The functions $F_0'(r,t_d)$, $F_1'(r,t_d)$, and $F_2'(r,t_d)$ are defined as
\begin{align}
    F_0'(r,t_d) = &  \frac{3r^3 (r^2-t_d^2)^{3/2}}{2\beta^2}  K_2 \left( \frac{\beta r}{2} \right) , \\
    F_1'(r,t_d) = & - \frac{r^3 \sqrt{r^2-t_d^2}}{\beta^2} \left[ 2\beta r(r^2-t_d^2) K_1 \left( \frac{\beta r}{2} \right) +3(r^2-5t_d^2) K_2 \left( \frac{\beta r}{2} \right) \right] , \\
    F_2'(r,t_d) = & \frac{r^2}{2\beta^3 \sqrt{r^2-t_d^2}} \left[ \beta r \{9r^4-90r^2t_d^2+105t_d^4+2\beta^2r^2(r^2-t_d^2)^2\} K_0 \left( \frac{\beta r}{2} \right) \right. \nonumber \\
    & \left. + 4 \{9r^4-90r^2t_d^2+105t_d^4+\beta^2r^2 (r^2-t_d^2)(r^2-5t_d^2) \} K_1 \left( \frac{\beta r}{2} \right) \right]  \, ,
\end{align}
which do not depend on $k$.

In contrast to the case of two-string loops,
there are no terms corresponding to constant-in-time sources 
because the non-overlapped regime $t_n > t_{12}$ does not contribute.
Consequently, $\Delta^\mr{single}(k/\beta)$ is not proportional to $k$ in the UV regime but becomes flat as shown below.
Note that 
each term in Eq.~\eqref{eq:Delta-single-result} has linear dependence on $k$.
However, there is cancellation in $\Delta^\mr{single}(k/\beta)$ to be the flat behavior $k^0$.
See Appendix~\ref{sec:k_cancel} for more details.

\subsection{Numerical result}

\begin{figure}[t]
    \centering
    \includegraphics[width=0.6\textwidth]{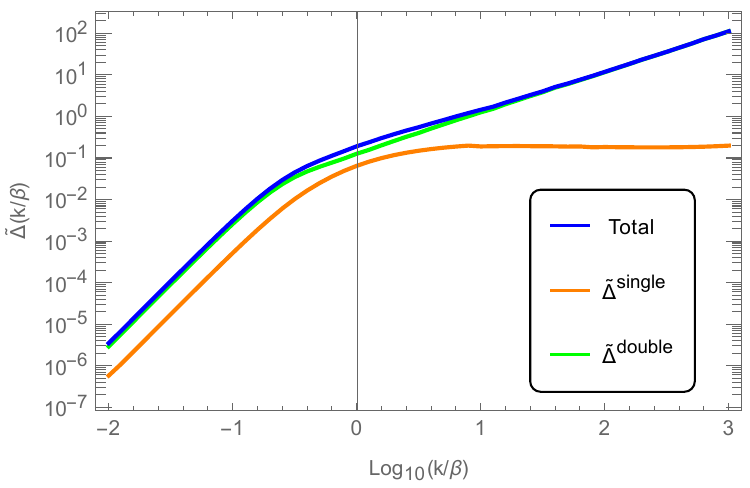}
    \caption{
    The numerical plot of $\tilde{\Delta}(k/\beta)=N_{\rm walls}^{-1}(H_*/\beta)^{-1} (\beta d_{\rm DW})^{-2}\Delta(k/\beta)$.
    The blue line is the sum of orange (single expanding string-loop contribution, Eq.~\eqref{eq:Delta-single-result}) and green (two expanding string-loop contribution, Eq.~\eqref{eq:delta_double}) lines.
    }
    \label{fig:delta}
\end{figure}  

Figure~\ref{fig:delta} shows the numerical plot of the total spectrum function $\Delta(k/\beta)$, Eq.~\eqref{eq:sum-Delta}.
For convention, it is normalized by 
a model-dependent factor $N_{\rm walls}^{-1}(H_*/\beta)^{-1} (\beta d_{\rm DW})^{-2}$.
One can see that the function exhibits asymptotic power laws,
\begin{align}
    \Delta^\mathrm{}(k/\beta) \propto
    \begin{cases}
        k^3 & (k/\beta\ll 1) \\
        k & (k/\beta\gg 1)
    \end{cases} \, .
\end{align}
In addition, $\Delta^\mathrm{}(k/\beta)$ can be fitted by the following function:
\begin{equation}
    \Delta^\mathrm{}(k/\beta) \simeq N_{\rm walls}\,\frac{H_*}{\beta} \, (\beta d_{\rm DW})^2
    \frac{a}{\left[ \left(k/\beta\right)^{-3/c}+b \left(k/\beta\right)^{-1/c}  \right]^c} ,
\end{equation}
with $a\simeq 2.75302$, $b\simeq 482.381$, and $c\simeq 0.51702$.
Note that one must have a UV cutoff for $k$, $k \sim \mathrm{min}\left(d_{\rm DW}^{-1}, R_c^{-1}\right)$ (see Sec.~\ref{sec:conclusion}),
above which our calculation is not reliable.
Also, $H_*/\beta < 1$ and $\beta d_{\rm DW} < 1$ work as suppression factors of the spectrum $\Delta(k/\beta)$.

The numerical plot shows that the contribution from two expanding string loops dominates the spectrum.
Moreover, the dominant contribution of two expanding string loops is coming from the term proportional to $g_{a,{\rm non}}''(r,t_d) g_{a,{\rm non}}''(r,-t_d)$, 
i.e., the both strings are nucleated in the non-overlapped regime, see Eq.~\eqref{eq:overlap-unoverlap}.
Also, this term gives a peculiar behavior that
$\Delta(k/\beta)$ linearly grows in the UV regime,
which is consistent with the argument in Ref.~\cite{RoperPol:2022iel}
since it provides a constant-source contribution in time,
as stated above.
Intuitively, the UV regime corresponds to small string loops.
Therefore, this implies that the small string loops radiate GWs even when they are perfectly circular before their collisions,
which is consistent with the analysis in Appendix~\ref{app-quadrupole}.
%

\section{Present stochastic GW spectrum}
\label{sec:present-spectrum}
We have seen so far the GW spectrum emitted from the hybrid defects of the DWs and cosmic strings.
Let us here consider how it would be observed by the current GW observatories.

\subsection{Production scenarios of hybrid defects}
The amplitude and typical frequency of the emitted GWs depend highly on how the DWs are produced and evolved.
We here present two representative scenarios; (A) scaling scenario and (B) non-thermal non-scaling scenario.
In both scenarios, we assume that the pre-existing string network is diluted away by the cosmic inflation.
For the scenario (A), 
the phase transition to produce the walls takes place with either of thermal or non-thermal effects.
Afterward, due to the free random motion of the walls and their reconnection,
the walls evolve into the scaling regime,
in which the number of DWs within one Hubble patch becomes of the order of unity.
Eventually, the string loops are nucleated and start to expand.
In this scenario, the spectrum is insensitive to the detailed way of the wall production as long as they reach the scaling regime.

For scenario (B), on the other hand,
the fields constituting the walls are decoupled from the thermal bath of the universe before the stage of the production of the walls.
Thus the fields feel Hubble friction, which prevents them from rolling down in the potential.
The phase transition to produce the walls takes place when the Hubble friction is balanced with the negative curvature of the potential.
The typical width of the created walls can be as large as the Hubble length scale at that time
while the curvature radius of the DWs depends on how the fluctuation of the field was produced; 
For instance, the random kick during the inflation produces scale-invariant fluctuation.
Here we do not specify the curvature radius since it is beyond the scope of this paper, but assume it to be larger than $1/\beta$ for viability of our calculation as stated above.
Subsequently, string loops are supposed to be nucleated well before the walls evolve into the scaling regime.
In this scenario, the GW spectrum can be amplified compared to scenario (A) because of two reasons:
the wall width $d_\mr{DW}$ (the square of which the spectrum function $\Delta(k)$ is proportional to) can be taken as close to the Hubble length $H_*$ at most,
and the non-scaling configuration can give a large number of walls $N_\mr{walls}$.

\subsection{Redshift}
The spectrum of the emitted GWs is affected by the redshift due to the Hubble expansion of the universe.
The scale factor at the phase transition (string loop creation) $a_*$ and at the present $a_0$ is related by
\begin{align}
    \frac{a_*}{a_0} = 7.97 \times 10^{-16} \left( \frac{g_*}{100} \right)^{-\frac{1}{3}} \left( \frac{T_*}{100 \, \mr{GeV}} \right)^{-1} , 
\end{align}
by using the entropy conservation during the radiation dominated era, $s_* a_*^3 = s_0 a_0^3$. 
Here, ``$*$'' denotes the time around when the loop creation completed and $g_*$ is the total number of the relativistic degrees of freedom at $T=T_*$.
The emitted frequency $f_*$ and the present red-shifted frequency $f_0$ are related as
\begin{align}
    f_0 = 1.65 \times 10^{-5} \ {\rm Hz} \left( \frac{f_*}{\beta} \right) \left( \frac{\beta}{H_*} \right) \left( \frac{T_*}{100  \ {\rm GeV}} \right) \left( \frac{g_*}{100 } \right)^{\frac{1}{6}} .
\end{align}
Here, $H_* = \sqrt{g_* \pi^2/90} \, T_*^2/M_{Pl}$, where $M_{Pl} = 1/\sqrt{8\pi G} \approx 2.435 \times 10^{18}$ GeV.
The energy fraction of the gravitational waves at present is defined as
\begin{align}
    (\Omega_{\rm GW} h^2 )_0 \equiv \frac{\rho_\mr{GW,0}}{\rho_c}h^2 = & 1.64 \times 10^{-5} \left( \frac{g_*}{100} \right)^{-\frac{1}{3}} \Omega_{\rm GW *} , \\
    = & 1.64 \times 10^{-5} \left( \frac{g_*}{100 } \right)^{-\frac{1}{3}} \kappa^2 \left( \frac{H_*}{\beta} \right)^2 \left( \frac{\alpha (t_*)}{1+\alpha (t_*)}  \right)^{2} \Delta (k_*) ,\label{eq:GWh2-0}
\end{align}
by using the energy conservation of the gravitational wave $(\Omega_{\rm GW} H^2 a^4 )_0 = (\Omega_{\rm GW} H^2 a^4 )_*$, $H_0 = 2.13 \times 10^{-42} h$ GeV and Eq.~\eqref{eq:relic2delta}.

\subsection{Present spectrum}
Because the nucleation processes are random and their typical length scale is sufficiently small compared to the current observation scale, 
the signal can be regarded as stochastic and almost isotropic.
Such a stochastic GW spectrum can be observed by the current and future GW observatories.

\subsubsection{Scaling solution case}
First, we consider case (A), where the DW became the scaling solution before the string loops are nucleated.
Here we take the width of the walls as $d_\mr{DW} \sim 1/v_\sigma$.
In this case, one should note that there is an additional contribution from the domain-wall network until the walls decay
since the network continuously generates GWs to remain in the scaling regime.
Such a spectrum is studied in Ref.~\cite{Dunsky:2021tih}
and found to be of the order of
\begin{equation}
\left. \left(\Omega_\mr{GW} h^2 \right)_0\right|_\text{wall network}\lesssim  10^{-4} \times \left(\frac{G \sigma}{H_*} \right)^2 \label{eq:GW-wall-network}
\end{equation}
at most.
On the other hand, in the scaling scenario, 
the amplitude of the GWs calculated in Eq.~\eqref{eq:GWh2-0} is estimated as
\begin{equation}
    (\Omega_{\rm GW} h^2 )_0 
\lesssim 10^{-6} \,  \frac{H_*}{v_\sigma} \left(\frac{\alpha(t_*)}{1+\alpha(t_*)}\right)^2 \, ,
\end{equation}
which approaches 
\begin{equation}
 \begin{cases} 
\displaystyle 10^{-6} \, \frac{H_*}{v_\sigma} & \alpha(t_*) \gg 1  \\[2ex]
\displaystyle  10^{-6} \, \frac{H_*}{v_\sigma} \left(\frac{G \sigma}{H_*}\right)^2 & \alpha(t_*) \ll 1  
 \end{cases} \, ,
\end{equation}
where we have substituted $k_\mr{max}\sim 1/d_\mr{DW}$
and have used $\alpha(t_*)\sim G \sigma/H_*$ at the scaling regime.
These expressions are much smaller than Eq.~\eqref{eq:GW-wall-network} in both regions $\alpha(t_*)\gtrless 1$.
Therefore, in scenario (A), 
the hybrid defects do not give a significant amount of the GWs,
and hence the main contribution to the GW spectrum comes from the wall network, Eq.~\eqref{eq:GW-wall-network}.\footnote{This may imply that the energy fraction parameter $x$ in Eq.~(95) of Ref.~\cite{Dunsky:2021tih} is quite small, $x\approx 0$.}

\subsubsection{Non-thermal and non-scaling case}

\begin{figure}[tbp]
    \centering    
    \includegraphics[width=0.8\textwidth]{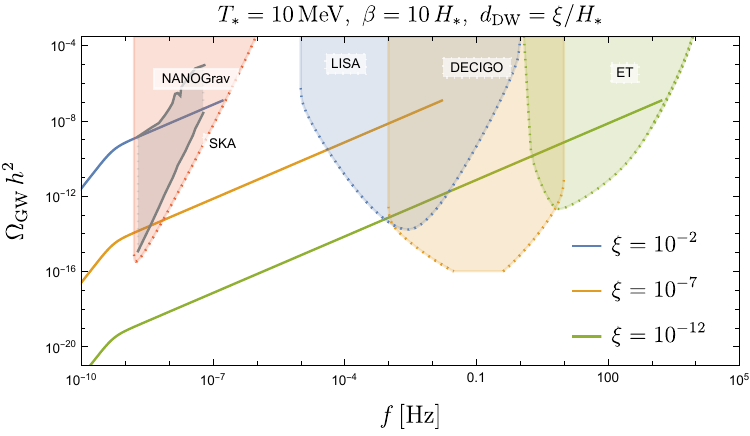}  \\[3em]
     \includegraphics[width=0.8\textwidth]{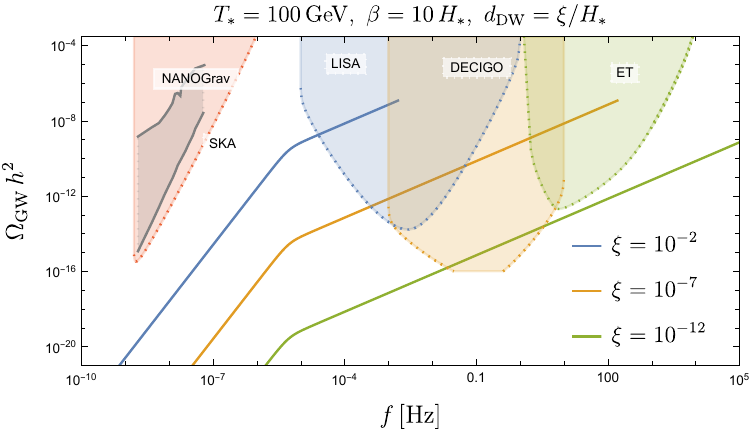}
    \caption{
    Present GW spectrum produced by expansion of string loops for the case of non-thermal production and non-scaling configuration of DWs.
    The string loops are assumed to be nucleated well before the DWs follow the scaling regime.
    Each curve corresponds to different choices of $\xi$, 
    which parametrizes the DW width (see the text).
    The DW VEV is taken so that $\alpha\sim 1$ for each line.
    The sensitivity curves of LISA, DECIGO, ET, and SKA are also shown.
    We also show the NANOGrav 15-year signal as the gray region.
    One can see that the blue curve in the upper panel can explain the NANOGrav signal. 
    For other parameters, we expect that they will be tested by future GW observatories.
    }
    \label{fig:spectrum_nonth}
\end{figure}

Next, we consider case (B), where the walls are produced non-thermally and never reach the scaling regime.
Because the typical width of the produced walls is around the Hubble length at the production stage,
we can take the width of the walls as $d_\mr{DW}\sim d_H|_\text{wall production}$,
which we parametrize as $d_\mr{DW}=\xi/H_*$ with $\xi \leq 1$.
Moreover, due to the non-scaling regime,
$N_\mr{walls}$ is not necessarily $\mathcal{O}(1)$, but is taken as $N_\mr{walls}=d_{H_*}/d_\mr{DW}$ as the most optimistic (densest) case.
If the DWs are immediately produced since the fields start to roll down in the potential
and the string loops are nucleated just after the wall production, 
then $\xi$ would be $\mathcal{O}(1)$.
Taking into account some time lag, we take some benchmark values of $\xi(<1)$.
The maximum value of the amplitude of the GWs in Eq.~\eqref{eq:GWh2-0} is estimated as
\begin{equation}
    (\Omega_{\rm GW} h^2 )_0 
\lesssim  10^{-6}  \left( \frac{H_*}{\beta} \right)^2 \left(\frac{\alpha(t_*)}{1+\alpha(t_*)}\right)^2 k_* \, d_\mr{DW} \, 
\end{equation}
from which one can expect significant amount of the GWs.

Figure.~\ref{fig:spectrum_nonth} shows the GW spectrum emitted from the hybrid defects 
with $T_*=10\, \mr{MeV}$ (top) and $T_*=100\, \mr{GeV}$ (bottom).
The wall tension is taken so that the energy fraction $\alpha(t_*)$ exceeds the order of unity,
$\sigma \sim (d_\mr{DW}/d_{H_*}) T_*^2 M_\mr{Pl}$.
We overlay the signal reported by the NANOGrav collaboration based on a 15-year data set~\cite{NANOGrav:2023gor} as the gray region
and project sensitivity curves of planned future observatories:
DECIGO~\cite{Kawamura:2020pcg} (orange), 
Einstein Telescope (ET)~\cite{Maggiore:2019uih} (green),
LISA~\cite{Bartolo:2016ami} (blue), 
and Square Kilometer Array (SKA)~\cite{Janssen:2014dka} (red). 
The sensitivity curves are taken from Ref.~\cite{Schmitz:2020syl}.
Note that we put the cutoff for the higher frequency region at $f_*=1/d_\mr{DW}$.
For other possible UV cutoffs, see Sec.~\ref{sec:conclusion}.
Remarkably, the NANOGrav signal may be explained by the GW spectrum from the hybrid defects when $T_*=10\, \mr{MeV}$, $\beta = 10H_*$, $d_\mr{DW}=10^{-2}/H_*$, and $\sigma\simeq (10^4\, \mr{GeV})^3$ (blue curve in the top panel).

Note that in this scenario (B),
the domain-wall network does not emit sufficient amount of GWs 
because the walls are not in the scaling regime.
Thus, the hybrid defects give the leading contribution.

\section{Discussion and conclusion}
\label{sec:conclusion}

In this paper, we analytically calculated the GW spectrum emitted from hybrid defects of DWs and cosmic strings, especially expanding string loops on the DWs.
The derived spectrum agrees with the universal behavior $f^3$ ($f$: frequency) in the IR regime below $\beta$,
whereas it becomes linear behavior in the UV regime, 
exhibiting a clear contrast to any other GW sources.
This is not counter-intuitive because the expanding loops can radiate GWs even when they are perfectly circular before the collisions.
Indeed, the circular-shape loop only has the $O(2)$ symmetry, 
which does not prevent the GW emission but has the non-zero and time-dependent quadrupole moment of the energy density.
Moreover, it turned out that the pulsar timing data in the $\mathcal{O}(1-10)$ nHz frequency band recently reported by NANOGrav, EPTA, PPTA, and CPTA groups
can be attributed to the predicted GW spectrum from the hybrid defects with appropriate parameters.
This GW spectrum is also detectable by future GW observatories
such as DECIGO, Einstein Telescope, LISA, and Square Kilometer Array.

Note that our calculation has three UV cutoffs,
corresponding to the initial radius of the nucleated loops $R_c$, the DW width $d_\mr{DW}$,
and the string width $d_\mr{st}$.
In the $f_* \gtrsim R_c^{-1}$ regime, 
we should take into account the finite size of the nucleated loops and 
modify the geometric consideration of the past light cones due to the finite speed of the expanding loops.
For $f_*\gtrsim d_\mr{st}^{-1}$, we have to take into account microscopic degrees of freedom of the strings,
which may modify the spectrum in such a regime.
For $f_*\gtrsim d_\mr{DW}^{-1}$, our thin-wall approximation breaks down.
In order for the nucleated loops to expand,
$R_c>d_\mr{st}$ is required by definition,
leading to the UV cutoff $f_* = \mr{min}(d_\mr{DW}^{-1}, R_c^{-1})$.
Since we considered the non-thermal scenario for the production of the DWs in order to provide significant amount of GWs,
$d_\mr{DW}$ is typically the largest length scale among them, so that we put the cutoff $f_* = d_\mr{DW}^{-1}$ in Fig.~\ref{fig:spectrum_nonth}.

While our result should be reliable as long as the assumptions we made are met,
it is still important to check the agreement in a different way, such as field-theoretic numerical simulations.
It is also important to calculate the macroscopic parameters from the field-theoretic point of view,
which would be helpful in order to distinguish/exclude models of particle physics.  
Furthermore, the envelope approximation that we rely on through the calculation 
might be eliminated 
using a similar strategy adopted in the bubble nucleation case~\cite{Jinno:2017fby},
which allows us to obtain a more precise result.
As an application, our calculation might be applied to decaying DWs triggered by primordial black holes~\cite{Stojkovic:2005zh},
which can be an alternative probe for them.

\section*{Acknowledgements}

The authors thank Simone Blasi and Ryusuke Jinno for useful discussions.
The authors also thank Thomas Konstandin for useful discussion and comments on the manuscript,
as well as Chiara Caprini for bringing a relevant paper to their attention.
This work is supported by JSPS Grant-in-Aid for
Scientific Research
KAKENHI Grant No.~JP22KJ3123 (Y.\ H.) and
Grant No.~JP23KJ2173 (W.\ N.),
and the Deutsche Forschungsgemeinschaft under Germany's Excellence Strategy - EXC 2121 Quantum Universe - 390833306.

\appendix

\section{GW from circular shape loop}
\label{app-quadrupole}
One may consider that the expanding uncollided string loop with the circular shape does not radiate GWs due to the symmetry.
However, this is not the case.
Let us calculate the quadrupole of the single loop explicitly.
The produced GWs in the TT gauge is proportional to the second time derivative of the quadrupole $Q_{ij}$ of the energy density,
\begin{align}
   h_{ij}^\mr{TT}& \propto (P_{ik} P_{jl} -\frac{1}{2} P_{ij}P_{kl}) \ddot{Q}^{kl} , \\
   Q_{ij}&=\int d^3 x \, \rho^s(x) \left[x_i x_j - \frac{1}{3} \delta_{ij} |\vec x|^2 \right] , \\
   P^{ij}&=\delta^{ij}-\hat{n}_{\rm obs}^i \hat{n}_{\rm obs}^j , 
\end{align}
where $\rho^s (x)$ is given by Eq.~\eqref{eq:rho-string}
and $\rho^s (x) = 0$ off the wall.
Here, $\hat{n}_{\rm obs}^i$ represents the unit vector pointing to an observer.
The DW is located on the $x-z$ plane.
Assuming that the hole is nucleated at $t=0$,
leading to $r_s(t) \simeq t$,
then we get 
\begin{align}
   Q_{ij}&= d_\mr{DW} \int_0^{2\pi} d \phi \int_t^{t+\ell_s} dr r \,
   \frac{\kappa \rho_\mr{re} t}{2 \ell_s} 
   \left[x_i x_j - \frac{1}{3} \delta_{ij} r^2 \right] \\
   &=\int_0^{2\pi} d \phi \,
   \frac{\kappa \rho_\mr{re}t }{8 \ell_s}   d_\mathrm{DW}   
\left[(t+\ell_s)^4 - t^4\right]
   \left[
\begin{pmatrix}
    \cos^2\phi  & 0 & \cos \phi \sin \phi \\ 
    0 & 0 & 0 \\
    \cos \phi \sin \phi & 0 & \sin^2 \phi 
\end{pmatrix}
- \frac{1}{3} \delta_{ij}
\right] \\
   &=
   \frac{\kappa \rho_\mr{re}}{2}  d_\mathrm{DW} t^4
\begin{pmatrix}
   \displaystyle +\frac{1}{6} & 0 & 0 \\ 
   0 & \displaystyle -\frac{1}{3} & 0 \\
   0 & 0 & \displaystyle +\frac{1}{6}
\end{pmatrix} \, .
\end{align}
The direction of the observer is set as $\hat{n}_{\rm obs} = (\sin\theta_{\rm obs} \cos\phi_{\rm obs},\cos\theta_{\rm obs},\sin\theta_{\rm obs} \sin\phi_{\rm obs})$.
The TT projection acts on the quadrupole as
\begin{align}
    h^{\rm TT}_{ij} \propto & \frac{\kappa \rho_\mr{re}}{2}  d_\mathrm{DW} t^2 \sin^2\theta_{\rm obs} 
    \begin{pmatrix}
   \displaystyle \tilde{h}_{11} & \tilde{h}_{12} & \tilde{h}_{13} \\ 
   \tilde{h}_{12} & \displaystyle \tilde{h}_{22} & \tilde{h}_{23} \\
   \tilde{h}_{13} & \tilde{h}_{23} & \displaystyle \tilde{h}_{33}
\end{pmatrix} \, ,
\end{align}
where
\begin{align}
    \tilde{h}_{11} = & \frac{3}{4}[1-3\cos(2\phi_{\rm obs})-2\cos(2\theta_{\rm obs})\cos^2\phi_{\rm obs}] , \\
    \tilde{h}_{12} = & 3\cos\theta_{\rm obs}\sin\theta_{\rm obs}\cos\phi_{\rm obs} , \\
    \tilde{h}_{13} = & -\frac{3}{4}\sin(2\phi_{\rm obs})[3+\cos(2\theta_{\rm obs})] , \\
    \tilde{h}_{22} = & -3\sin^2\theta_{\rm obs} , \\
    \tilde{h}_{23} = & 3\cos\theta_{\rm obs}\sin\theta_{\rm obs}\sin\phi_{\rm obs} , \\
    \tilde{h}_{33} = & \frac{3}{4}[1+3\cos(2\phi_{\rm obs})-2\cos(2\theta_{\rm obs})\sin^2\phi_{\rm obs}] .
\end{align}
Therefore, $h^{\rm TT}_{ij} \neq 0$, and hence the circular-shape loop does radiate GWs.

From the expression of $Q_{ij}$, one can roughly deduce the GW spectrum $\Delta(k/\beta)$ radiated from expanding loops based on a simple model.
Let us assume that all of the string loops have roughly the same radius $r_s(t)$.
The radiated GW energy per unit time is expressed with the quadrupole as
\begin{align}
 \frac{d E_\mr{GW}}{dt} = \frac{G}{5} \left(\dddot{Q}_{ij}\right)^2 \, .
\end{align}
Since the time scale of the decay process of the DWs is given by $\beta^{-1}$,
the GW energy radiated from each loop during the process is estimated as
\begin{align}
 \Delta E_\mr{GW} \sim \frac{G}{\beta} \left(\kappa \rho_{re} \,d_\mr{DW} \,t \right)^2 \, .
\end{align}
Summing this over multiple loops on the DW, 
the total GW energy is given as
\begin{align}
\sum_\text{loops} \Delta E_\mr{GW} &\sim G \sum_{a,b}  \dddot{Q}_{ij}^a \dddot{Q}_{ij}^b \\
&\sim G \sum_{a}  \left(\dddot{Q}_{ij}^a\right)^2 + G \sum_{a\neq b} \dddot{Q}_{ij}^a \dddot{Q}_{ij}^b
\end{align}
where $a,b$ label the string loops.
In the first term, the summation gives a factor $\sim 1/(\beta r_s(t))^2$,
corresponding to the number of loops within the relevant system area $1/\beta^2$.
In the second term, the summation gives the number of pairs of the loops that are nucleated within the time interval $r_s(t)$
(otherwise they cannot communicate with each other causally),
which is given as 
\begin{equation}
\frac{1}{(\beta r_s(t))^4}  \beta r_s(t) \sim \frac{1}{(\beta r_s(t))^3} \, .
\end{equation}
Noting $r_s(t)\sim t$, we get
\begin{align}
\sum_\text{loops} \Delta E_\mr{GW} 
\sim \frac{G}{\beta^3} \left(\kappa \rho_{re} \,d_\mr{DW} \right)^2
\left[1 + \frac{1}{\beta t} \right] \, .
\end{align}
Translating this into the spectrum function $\Delta$ with the loop radius $r_s\sim t$ identified with the inverse of the frequency $1/k$, 
roughly we get
\begin{align}
 \Delta(k) \sim \frac{H_*}{\beta} (\beta \, d_\mr{DW})^2 \left[1+\frac{k}{\beta}\right]
\end{align}
per one DW.
The first term does not depend on the loop radius and agrees with the flat UV behavior of $\Delta^\mr{single}(k)$ in Eq.~\eqref{eq:Delta-single-result} and in Fig.~\ref{fig:delta},
while the second term is proportional to $k$ linearly and agrees with $\Delta^\mr{double}(k)$ in the UV regime.

\section{(Next) Leading $k$-cancellation}
\label{sec:k_cancel}

\begin{figure}[t]
    \centering
    \includegraphics[width=0.5\textwidth]{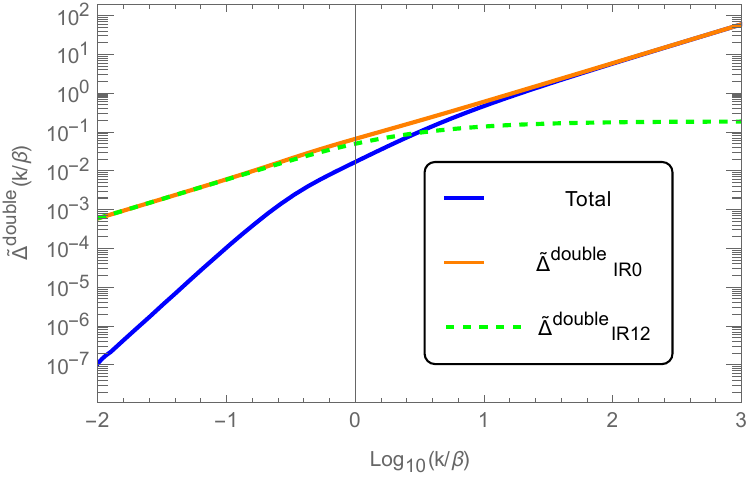}
    \caption{
    The spectrum $\tilde{\Delta}^{\rm double}(k/\beta)=N_{\rm walls}^{-1}(H_*/\beta)^{-1} (\beta d_{\rm DW})^{-2}\Delta^{\rm double}(k/\beta)$ of Eq.~\eqref{eq:ddc12} (green line) and Eq.~\eqref{eq:ddc0n} (orange line).
    The blue line is the sum of Eqs.~\eqref{eq:ddc12} and \eqref{eq:ddc0n}. 
    The positive value is shown by the solid line. 
    For the negative value, its absolute value is shown by dashed line. 
    }
    \label{fig:delta_double_const}
\end{figure}  

\begin{figure}[t]
    \centering
    \includegraphics[width=0.5\textwidth]{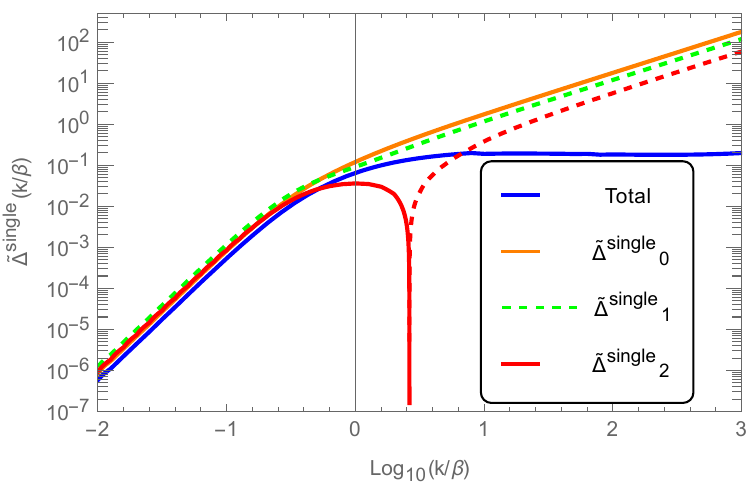}
    \caption{
    The quantity $\tilde{\Delta}^{\rm single}(k/\beta)=N_{\rm walls}^{-1} (H_*/\beta)^{-1} (\beta d_{\rm DW})^{-2} \Delta^{\rm single}(k/\beta)$ is calculated by summing up each $j_i(kr)$ component (blue line).
    The positive value is shown by the solid lines.  
    The negative value is taken its absolute value and shown by dashed line.
    The sign of $\Delta^{\rm single}_2(k/\beta)$ is positive for $\log_{10} (k/\beta) \lesssim 0.4$ and negative for $\log_{10} (k/\beta) \gtrsim 0.4$.
    The linear $k$ dependence of $\Delta^{\rm single}_i(k/\beta)$ is canceled by summing up.
    }
    \label{fig:delta_single_log}
\end{figure}

Here, the numerical plots of Eqs.~\eqref{eq:ddc12} and \eqref{eq:ddc0n} are shown in Fig.~\ref{fig:delta_double_const}.
The result shows that the dependence of $k^{1}$ (and $k^{2}$) is canceled in the small $k$ region.
Similarly, the numerical plots of $\Delta^{\rm single}(k/\beta)$ are shown in Fig.~\ref{fig:delta_single_log}.
Here, the spectrum function $\Delta^{\rm single}(k/\beta)$ is decomposed as
\begin{align}
    \Delta^{\rm single}_0 (k/\beta)= & \frac{3}{4\pi} \beta k^3 \int_{0}^{\infty} dt_d \cos (k t_d) \int_{t_d}^{\infty} dr \frac{C_{\rm DW}(r)}{r^3 \mathcal{I}(r,t_d)}  j_0 (kr) F_0''(r,t_d)  , \\
    \Delta^{\rm single}_1(k/\beta) = & \frac{3}{4\pi} \beta k^3 \int_{0}^{\infty} dt_d \cos (k t_d) \int_{t_d}^{\infty} dr \frac{C_{\rm DW}(r)}{r^3 \mathcal{I}(r,t_d)}   \frac{j_1(kr)}{kr} F_1''(r,t_d)  , \\
    \Delta^{\rm single}_2(k/\beta) = & \frac{3}{4\pi} \beta k^3 \int_{0}^{\infty} dt_d \cos (k t_d) \int_{t_d}^{\infty} dr \frac{C_{\rm DW}(r)}{r^3 \mathcal{I}(r,t_d)} \frac{j_2(kr)}{(kr)^2} F_2''(r,t_d)  .
\end{align}
Then, the linear $k$ dependence is canceled in the large $k$ region in this case.\footnote{
For the case of single bubble nucleation in FOPT,
the GW spectrum apparently has $k$ dependence as $k^0$ in the UV regime,
which is actually canceled and leads to a resultant leading-order contribution as $k^{-1}$.
Also, for the case of double bubble nucleation in FOPT, 
the apparent $k^{-1}$ dependence vanishes and $k^{-2}$ dependence remains.
}

\bibliographystyle{jhep}

\bibliography{references}

\providecommand{\href}[2]{#2}\begingroup\raggedright\begin{thebibliography}{10}

\bibitem{LIGOScientific:2016aoc}
{\scshape LIGO Scientific, Virgo} collaboration, \emph{{Observation of
  Gravitational Waves from a Binary Black Hole Merger}},
  \href{https://doi.org/10.1103/PhysRevLett.116.061102}{\emph{Phys. Rev. Lett.}
  {\bfseries 116} (2016) 061102}
  [\href{https://arxiv.org/abs/1602.03837}{{\ttfamily 1602.03837}}].

\bibitem{LIGOScientific:2016emj}
{\scshape LIGO Scientific, Virgo} collaboration, \emph{{GW150914: The Advanced
  LIGO Detectors in the Era of First Discoveries}},
  \href{https://doi.org/10.1103/PhysRevLett.116.131103}{\emph{Phys. Rev. Lett.}
  {\bfseries 116} (2016) 131103}
  [\href{https://arxiv.org/abs/1602.03838}{{\ttfamily 1602.03838}}].

\bibitem{LIGOScientific:2016vbw}
{\scshape LIGO Scientific, Virgo} collaboration, \emph{{GW150914: First results
  from the search for binary black hole coalescence with Advanced LIGO}},
  \href{https://doi.org/10.1103/PhysRevD.93.122003}{\emph{Phys. Rev. D}
  {\bfseries 93} (2016) 122003}
  [\href{https://arxiv.org/abs/1602.03839}{{\ttfamily 1602.03839}}].

\bibitem{LIGOScientific:2016vlm}
{\scshape LIGO Scientific, Virgo} collaboration, \emph{{Properties of the
  Binary Black Hole Merger GW150914}},
  \href{https://doi.org/10.1103/PhysRevLett.116.241102}{\emph{Phys. Rev. Lett.}
  {\bfseries 116} (2016) 241102}
  [\href{https://arxiv.org/abs/1602.03840}{{\ttfamily 1602.03840}}].

\bibitem{LIGOScientific:2017vwq}
{\scshape LIGO Scientific, Virgo} collaboration, \emph{{GW170817: Observation
  of Gravitational Waves from a Binary Neutron Star Inspiral}},
  \href{https://doi.org/10.1103/PhysRevLett.119.161101}{\emph{Phys. Rev. Lett.}
  {\bfseries 119} (2017) 161101}
  [\href{https://arxiv.org/abs/1710.05832}{{\ttfamily 1710.05832}}].

\bibitem{Kosowsky:1991ua}
A.~Kosowsky, M.~S. Turner and R.~Watkins, \emph{{Gravitational radiation from
  colliding vacuum bubbles}},
  \href{https://doi.org/10.1103/PhysRevD.45.4514}{\emph{Phys. Rev. D}
  {\bfseries 45} (1992) 4514}.

\bibitem{Kosowsky:1992vn}
A.~Kosowsky and M.~S. Turner, \emph{{Gravitational radiation from colliding
  vacuum bubbles: envelope approximation to many bubble collisions}},
  \href{https://doi.org/10.1103/PhysRevD.47.4372}{\emph{Phys. Rev. D}
  {\bfseries 47} (1993) 4372}
  [\href{https://arxiv.org/abs/astro-ph/9211004}{{\ttfamily
  astro-ph/9211004}}].

\bibitem{Kamionkowski:1993fg}
M.~Kamionkowski, A.~Kosowsky and M.~S. Turner, \emph{{Gravitational radiation
  from first order phase transitions}},
  \href{https://doi.org/10.1103/PhysRevD.49.2837}{\emph{Phys. Rev. D}
  {\bfseries 49} (1994) 2837}
  [\href{https://arxiv.org/abs/astro-ph/9310044}{{\ttfamily
  astro-ph/9310044}}].

\bibitem{Grojean:2006bp}
C.~Grojean and G.~Servant, \emph{{Gravitational Waves from Phase Transitions at
  the Electroweak Scale and Beyond}},
  \href{https://doi.org/10.1103/PhysRevD.75.043507}{\emph{Phys. Rev. D}
  {\bfseries 75} (2007) 043507}
  [\href{https://arxiv.org/abs/hep-ph/0607107}{{\ttfamily hep-ph/0607107}}].

\bibitem{Caprini:2009fx}
C.~Caprini, R.~Durrer, T.~Konstandin and G.~Servant, \emph{{General Properties
  of the Gravitational Wave Spectrum from Phase Transitions}},
  \href{https://doi.org/10.1103/PhysRevD.79.083519}{\emph{Phys. Rev. D}
  {\bfseries 79} (2009) 083519}
  [\href{https://arxiv.org/abs/0901.1661}{{\ttfamily 0901.1661}}].

\bibitem{Caprini:2015zlo}
C.~Caprini et~al., \emph{{Science with the space-based interferometer eLISA.
  II: Gravitational waves from cosmological phase transitions}},
  \href{https://doi.org/10.1088/1475-7516/2016/04/001}{\emph{JCAP} {\bfseries
  04} (2016) 001} [\href{https://arxiv.org/abs/1512.06239}{{\ttfamily
  1512.06239}}].

\bibitem{Caprini:2018mtu}
C.~Caprini and D.~G. Figueroa, \emph{{Cosmological Backgrounds of Gravitational
  Waves}}, \href{https://doi.org/10.1088/1361-6382/aac608}{\emph{Class. Quant.
  Grav.} {\bfseries 35} (2018) 163001}
  [\href{https://arxiv.org/abs/1801.04268}{{\ttfamily 1801.04268}}].

\bibitem{Caprini:2019egz}
C.~Caprini et~al., \emph{{Detecting gravitational waves from cosmological phase
  transitions with LISA: an update}},
  \href{https://doi.org/10.1088/1475-7516/2020/03/024}{\emph{JCAP} {\bfseries
  03} (2020) 024} [\href{https://arxiv.org/abs/1910.13125}{{\ttfamily
  1910.13125}}].

\bibitem{Athron:2023xlk}
P.~Athron, C.~Bal\'azs, A.~Fowlie, L.~Morris and L.~Wu, \emph{{Cosmological
  phase transitions: From perturbative particle physics to gravitational
  waves}}, \href{https://doi.org/10.1016/j.ppnp.2023.104094}{\emph{Prog. Part.
  Nucl. Phys.} {\bfseries 135} (2024) 104094}
  [\href{https://arxiv.org/abs/2305.02357}{{\ttfamily 2305.02357}}].

\bibitem{Vilenkin:1981bx}
A.~Vilenkin, \emph{{Gravitational radiation from cosmic strings}},
  \href{https://doi.org/10.1016/0370-2693(81)91144-8}{\emph{Phys. Lett. B}
  {\bfseries 107} (1981) 47}.

\bibitem{Accetta:1988bg}
F.~S. Accetta and L.~M. Krauss, \emph{{The stochastic gravitational wave
  spectrum resulting from cosmic string evolution}},
  \href{https://doi.org/10.1016/0550-3213(89)90628-7}{\emph{Nucl. Phys. B}
  {\bfseries 319} (1989) 747}.

\bibitem{Caldwell:1991jj}
R.~R. Caldwell and B.~Allen, \emph{{Cosmological constraints on cosmic string
  gravitational radiation}},
  \href{https://doi.org/10.1103/PhysRevD.45.3447}{\emph{Phys. Rev. D}
  {\bfseries 45} (1992) 3447}.

\bibitem{Vilenkin:2000jqa}
A.~Vilenkin and E.~S. Shellard, \emph{{Cosmic Strings and Other Topological
  Defects}}. Cambridge University Press, 7, 2000.

\bibitem{Kibble:1982dd}
T.~W.~B. Kibble, G.~Lazarides and Q.~Shafi, \emph{{Walls Bounded by Strings}},
  \href{https://doi.org/10.1103/PhysRevD.26.435}{\emph{Phys. Rev. D} {\bfseries
  26} (1982) 435}.

\bibitem{Vilenkin:1982ks}
A.~Vilenkin and A.~E. Everett, \emph{{Cosmic Strings and Domain Walls in Models
  with Goldstone and PseudoGoldstone Bosons}},
  \href{https://doi.org/10.1103/PhysRevLett.48.1867}{\emph{Phys. Rev. Lett.}
  {\bfseries 48} (1982) 1867}.

\bibitem{Everett:1982nm}
A.~E. Everett and A.~Vilenkin, \emph{{Left-right Symmetric Theories and Vacuum
  Domain Walls and Strings}},
  \href{https://doi.org/10.1016/0550-3213(82)90135-3}{\emph{Nucl. Phys. B}
  {\bfseries 207} (1982) 43}.

\bibitem{Preskill:1992ck}
J.~Preskill and A.~Vilenkin, \emph{{Decay of metastable topological defects}},
  \href{https://doi.org/10.1103/PhysRevD.47.2324}{\emph{Phys. Rev. D}
  {\bfseries 47} (1993) 2324}
  [\href{https://arxiv.org/abs/hep-ph/9209210}{{\ttfamily hep-ph/9209210}}].

\bibitem{Kawasaki:2013ae}
M.~Kawasaki and K.~Nakayama, \emph{{Axions: Theory and Cosmological Role}},
  \href{https://doi.org/10.1146/annurev-nucl-102212-170536}{\emph{Ann. Rev.
  Nucl. Part. Sci.} {\bfseries 63} (2013) 69}
  [\href{https://arxiv.org/abs/1301.1123}{{\ttfamily 1301.1123}}].

\bibitem{Eto:2023gfn}
M.~Eto, Y.~Hamada and M.~Nitta, \emph{{Composite topological solitons
  consisting of domain walls, strings, and monopoles in O(N) models}},
  \href{https://doi.org/10.1007/JHEP08(2023)150}{\emph{JHEP} {\bfseries 08}
  (2023) 150} [\href{https://arxiv.org/abs/2304.14143}{{\ttfamily
  2304.14143}}].

\bibitem{Maji:2023fba}
R.~Maji, W.-I. Park and Q.~Shafi, \emph{{Gravitational waves from walls bounded
  by strings in SO(10) model of pseudo-Goldstone dark matter}},
  \href{https://doi.org/10.1016/j.physletb.2023.138127}{\emph{Phys. Lett. B}
  {\bfseries 845} (2023) 138127}
  [\href{https://arxiv.org/abs/2305.11775}{{\ttfamily 2305.11775}}].

\bibitem{Lazarides:2023ksx}
G.~Lazarides, R.~Maji and Q.~Shafi, \emph{{Superheavy quasistable strings and
  walls bounded by strings in the light of NANOGrav 15~year data}},
  \href{https://doi.org/10.1103/PhysRevD.108.095041}{\emph{Phys. Rev. D}
  {\bfseries 108} (2023) 095041}
  [\href{https://arxiv.org/abs/2306.17788}{{\ttfamily 2306.17788}}].

\bibitem{Fu:2024rsm}
B.~Fu, A.~Ghoshal, S.~F. King and M.~H. Rahat, \emph{{Type-I two-Higgs-doublet
  model and gravitational waves from domain walls bounded by strings}},
  \href{https://arxiv.org/abs/2404.16931}{{\ttfamily 2404.16931}}.

\bibitem{Dunsky:2021tih}
D.~I. Dunsky, A.~Ghoshal, H.~Murayama, Y.~Sakakihara and G.~White, \emph{{GUTs,
  hybrid topological defects, and gravitational waves}},
  \href{https://doi.org/10.1103/PhysRevD.106.075030}{\emph{Phys. Rev. D}
  {\bfseries 106} (2022) 075030}
  [\href{https://arxiv.org/abs/2111.08750}{{\ttfamily 2111.08750}}].

\bibitem{Gleiser:1998na}
M.~Gleiser and R.~Roberts, \emph{{Gravitational waves from collapsing vacuum
  domains}}, \href{https://doi.org/10.1103/PhysRevLett.81.5497}{\emph{Phys.
  Rev. Lett.} {\bfseries 81} (1998) 5497}
  [\href{https://arxiv.org/abs/astro-ph/9807260}{{\ttfamily
  astro-ph/9807260}}].

\bibitem{Preskill:1991kd}
J.~Preskill, S.~P. Trivedi, F.~Wilczek and M.~B. Wise, \emph{{Cosmology and
  broken discrete symmetry}},
  \href{https://doi.org/10.1016/0550-3213(91)90241-O}{\emph{Nucl. Phys. B}
  {\bfseries 363} (1991) 207}.

\bibitem{Hiramatsu:2013qaa}
T.~Hiramatsu, M.~Kawasaki and K.~Saikawa, \emph{{On the estimation of
  gravitational wave spectrum from cosmic domain walls}},
  \href{https://doi.org/10.1088/1475-7516/2014/02/031}{\emph{JCAP} {\bfseries
  02} (2014) 031} [\href{https://arxiv.org/abs/1309.5001}{{\ttfamily
  1309.5001}}].

\bibitem{Saikawa:2017hiv}
K.~Saikawa, \emph{{A review of gravitational waves from cosmic domain walls}},
  \href{https://doi.org/10.3390/universe3020040}{\emph{Universe} {\bfseries 3}
  (2017) 40} [\href{https://arxiv.org/abs/1703.02576}{{\ttfamily 1703.02576}}].

\bibitem{Kitajima:2023cek}
N.~Kitajima, J.~Lee, K.~Murai, F.~Takahashi and W.~Yin, \emph{{Gravitational
  waves from domain wall collapse, and application to nanohertz signals with
  QCD-coupled axions}},
  \href{https://doi.org/10.1016/j.physletb.2024.138586}{\emph{Phys. Lett. B}
  {\bfseries 851} (2024) 138586}
  [\href{https://arxiv.org/abs/2306.17146}{{\ttfamily 2306.17146}}].

\bibitem{Jinno:2016vai}
R.~Jinno and M.~Takimoto, \emph{{Gravitational waves from bubble collisions: An
  analytic derivation}},
  \href{https://doi.org/10.1103/PhysRevD.95.024009}{\emph{Phys. Rev. D}
  {\bfseries 95} (2017) 024009}
  [\href{https://arxiv.org/abs/1605.01403}{{\ttfamily 1605.01403}}].

\bibitem{Jinno:2017fby}
R.~Jinno and M.~Takimoto, \emph{{Gravitational waves from bubble dynamics:
  Beyond the Envelope}},
  \href{https://doi.org/10.1088/1475-7516/2019/01/060}{\emph{JCAP} {\bfseries
  01} (2019) 060} [\href{https://arxiv.org/abs/1707.03111}{{\ttfamily
  1707.03111}}].

\bibitem{Blasi:2023rqi}
S.~Blasi, R.~Jinno, T.~Konstandin, H.~Rubira and I.~Stomberg,
  \emph{{Gravitational waves from defect-driven phase transitions: domain
  walls}}, \href{https://doi.org/10.1088/1475-7516/2023/10/051}{\emph{JCAP}
  {\bfseries 10} (2023) 051}
  [\href{https://arxiv.org/abs/2302.06952}{{\ttfamily 2302.06952}}].

\bibitem{Blasi:2024mtc}
S.~Blasi and A.~Mariotti, \emph{{QCD Axion Strings or Seeds?}},
  \href{https://arxiv.org/abs/2405.08060}{{\ttfamily 2405.08060}}.

\bibitem{NANOGrav:2023gor}
{\scshape NANOGrav} collaboration, \emph{{The NANOGrav 15 yr Data Set: Evidence
  for a Gravitational-wave Background}},
  \href{https://doi.org/10.3847/2041-8213/acdac6}{\emph{Astrophys. J. Lett.}
  {\bfseries 951} (2023) L8}
  [\href{https://arxiv.org/abs/2306.16213}{{\ttfamily 2306.16213}}].

\bibitem{EPTA:2023fyk}
{\scshape EPTA, InPTA:} collaboration, \emph{{The second data release from the
  European Pulsar Timing Array - III. Search for gravitational wave signals}},
  \href{https://doi.org/10.1051/0004-6361/202346844}{\emph{Astron. Astrophys.}
  {\bfseries 678} (2023) A50}
  [\href{https://arxiv.org/abs/2306.16214}{{\ttfamily 2306.16214}}].

\bibitem{Xu:2023wog}
H.~Xu et~al., \emph{{Searching for the Nano-Hertz Stochastic Gravitational Wave
  Background with the Chinese Pulsar Timing Array Data Release I}},
  \href{https://doi.org/10.1088/1674-4527/acdfa5}{\emph{Res. Astron.
  Astrophys.} {\bfseries 23} (2023) 075024}
  [\href{https://arxiv.org/abs/2306.16216}{{\ttfamily 2306.16216}}].

\bibitem{Reardon:2023gzh}
D.~J. Reardon et~al., \emph{{Search for an Isotropic Gravitational-wave
  Background with the Parkes Pulsar Timing Array}},
  \href{https://doi.org/10.3847/2041-8213/acdd02}{\emph{Astrophys. J. Lett.}
  {\bfseries 951} (2023) L6}
  [\href{https://arxiv.org/abs/2306.16215}{{\ttfamily 2306.16215}}].

\bibitem{PhysRevLett.55.2398}
M.~Hindmarsh and T.~W.~B. Kibble, \emph{Monopoles on strings},
  \href{https://doi.org/10.1103/PhysRevLett.55.2398}{\emph{Phys. Rev. Lett.}
  {\bfseries 55} (1985) 2398}.

\bibitem{Caprini:2007xq}
C.~Caprini, R.~Durrer and G.~Servant, \emph{{Gravitational wave generation from
  bubble collisions in first-order phase transitions: An analytic approach}},
  \href{https://doi.org/10.1103/PhysRevD.77.124015}{\emph{Phys. Rev. D}
  {\bfseries 77} (2008) 124015}
  [\href{https://arxiv.org/abs/0711.2593}{{\ttfamily 0711.2593}}].

\bibitem{Myers:1991yh}
E.~Myers, C.~Rebbi and R.~Strilka, \emph{{A Study of the interaction and
  scattering of vortices in the Abelian Higgs (or Ginzburg-Landau) model}},
  \href{https://doi.org/10.1103/PhysRevD.45.1355}{\emph{Phys. Rev. D}
  {\bfseries 45} (1992) 1355}.

\bibitem{RoperPol:2022iel}
A.~Roper~Pol, C.~Caprini, A.~Neronov and D.~Semikoz, \emph{{Gravitational wave
  signal from primordial magnetic fields in the Pulsar Timing Array frequency
  band}}, \href{https://doi.org/10.1103/PhysRevD.105.123502}{\emph{Phys. Rev.
  D} {\bfseries 105} (2022) 123502}
  [\href{https://arxiv.org/abs/2201.05630}{{\ttfamily 2201.05630}}].

\bibitem{Kawamura:2020pcg}
S.~Kawamura et~al., \emph{{Current status of space gravitational wave antenna
  DECIGO and B-DECIGO}},
  \href{https://doi.org/10.1093/ptep/ptab019}{\emph{PTEP} {\bfseries 2021}
  (2021) 05A105} [\href{https://arxiv.org/abs/2006.13545}{{\ttfamily
  2006.13545}}].

\bibitem{Maggiore:2019uih}
M.~Maggiore et~al., \emph{{Science Case for the Einstein Telescope}},
  \href{https://doi.org/10.1088/1475-7516/2020/03/050}{\emph{JCAP} {\bfseries
  03} (2020) 050} [\href{https://arxiv.org/abs/1912.02622}{{\ttfamily
  1912.02622}}].

\bibitem{Bartolo:2016ami}
N.~Bartolo et~al., \emph{{Science with the space-based interferometer LISA. IV:
  Probing inflation with gravitational waves}},
  \href{https://doi.org/10.1088/1475-7516/2016/12/026}{\emph{JCAP} {\bfseries
  12} (2016) 026} [\href{https://arxiv.org/abs/1610.06481}{{\ttfamily
  1610.06481}}].

\bibitem{Janssen:2014dka}
G.~Janssen et~al., \emph{{Gravitational wave astronomy with the SKA}},
  \href{https://doi.org/10.22323/1.215.0037}{\emph{PoS} {\bfseries AASKA14}
  (2015) 037} [\href{https://arxiv.org/abs/1501.00127}{{\ttfamily
  1501.00127}}].

\bibitem{Schmitz:2020syl}
K.~Schmitz, \emph{{New Sensitivity Curves for Gravitational-Wave Signals from
  Cosmological Phase Transitions}},
  \href{https://doi.org/10.1007/JHEP01(2021)097}{\emph{JHEP} {\bfseries 01}
  (2021) 097} [\href{https://arxiv.org/abs/2002.04615}{{\ttfamily
  2002.04615}}].

\bibitem{Stojkovic:2005zh}
D.~Stojkovic, K.~Freese and G.~D. Starkman, \emph{{Holes in the walls:
  Primordial black holes as a solution to the cosmological domain wall
  problem}}, \href{https://doi.org/10.1103/PhysRevD.72.045012}{\emph{Phys. Rev.
  D} {\bfseries 72} (2005) 045012}
  [\href{https://arxiv.org/abs/hep-ph/0505026}{{\ttfamily hep-ph/0505026}}].

\end{thebibliography}\endgroup
\end{document}